\newif\ifC
\newif\ifAnon
\Anontrue  

\ifC
  \ifAnon
    \documentclass[a4paper,USenglish,cleveref,autoref,thm-restate,anonymous]{socg-lipics-v2021}
  \else
    \documentclass[a4paper,USenglish,cleveref,autoref,thm-restate]{socg-lipics-v2021}
  \fi
\else
    \documentclass[11pt]{article}
\fi

\ifC
\else
\fi

\title{Near-Linear Almost Exact Diameter in Weighted Planar Digraphs}

\ifC
\titlerunning{Near-Linear Almost Exact Diameter in Weighted Planar Digraphs}

\author{Shinwoo An}
{Department of Computer Science, Bar-Ilan University, Ramat Gan 5290002, Israel}
{shinwoo.an@biu.ac.il}
{https://orcid.org/0009-0007-5671-6074}
{}

\author{Avi Kadria}
{Department of Computer Science, Bar-Ilan University, Ramat Gan 5290002, Israel}
{avi.kadria3@gmail.com}
{https://orcid.org/0000-0001-8449-3284}
{}

\author{Wolfgang Mulzer}
{Institut f\"ur Informatik, Freie Universit\"at Berlin, Germany}
{mulzer@inf.fu-berlin.de}
{https://orcid.org/0000-0002-1948-5840}
{}

\author{Liam Roditty}
{Department of Computer Science, Bar-Ilan University, Ramat Gan 5290002, Israel}
{liam.roditty@biu.ac.il}
{https://orcid.org/0000-0002-5289-198X}
{}

\authorrunning{S. An, A. Kadria, W. Mulzer, and L. Roditty}
\Copyright{Shinwoo An, Avi Kadria, Wolfgang Mulzer, and Liam Roditty}

\ccsdesc[500]{Theory of computation~Shortest paths}
\ccsdesc[500]{Theory of computation~Computational geometry}
\ccsdesc[300]{Theory of computation~Approximation algorithms analysis}
\ccsdesc[300]{Mathematics of computing~Graph algorithms}

\keywords{Planar graphs, directed graphs, diameter, eccentricities, approximation algorithms, shortest-path separators, orthogonal range searching}

\EventEditors{}
\EventNoEds{1}
\EventLongTitle{43rd International Symposium on Computational Geometry (SoCG 2027)}
\EventShortTitle{SoCG 2027}
\EventAcronym{SoCG}
\EventYear{2027}
\EventDate{June 21--25, 2027}
\EventLocation{Bengaluru, India}
\EventLogo{}
\SeriesVolume{0}
\ArticleNo{0}
\ifAnon\hideLIPIcs\fi
\else
\author{Shinwoo An \and Avi Kadria \and Wolfgang Mulzer \and Liam Roditty}
\date{}
\fi

\ifC\else
\usepackage[margin=1in]{geometry}
\usepackage{setspace}
\usepackage{cite}
\usepackage{fullpage}
\usepackage{enumitem}
\usepackage[bottom]{footmisc}
\fi
\usepackage{multirow}
\usepackage{makecell}
\usepackage{booktabs}

\usepackage{graphicx}
\usepackage{adjustbox}
\usepackage{tikz}
\usetikzlibrary{arrows,automata,patterns,decorations.pathreplacing}

\usepackage[linesnumbered,ruled,vlined,boxed,algo2e]{algorithm2e}
\usepackage{nicefrac}
\usepackage{amsmath}
\usepackage{float}
\usepackage{mathtools}
\usepackage{amsthm}
\usepackage{amsfonts}
\usepackage{clipboard}
\usepackage{hyperref}
\ifC
\newtheorem{question}{Question}
\else
\usepackage[nameinlink]{cleveref}
\newtheorem{theorem}{Theorem}

\newtheorem{lemma}{Lemma}

\newtheorem{corollary}{Corollary}[theorem]

\newtheorem{question}{Question}
\fi

\Crefname{enumi}{(item)}{(items)}

\newcommand{\Reminder}[1]{

\vspace{0.5em}
\noindent\textbf{Reminder of~\Cref{#1}.} \textit{\Paste{#1}}
\vspace{0.5em}

}

\usepackage{xcolor}
\definecolor{DarkGreen}{RGB}{1,50,32}
\usepackage{pgfplots}
\pgfplotsset{compat=1.18}

\usepackage{silence}
\ifC\else
\usepackage{chngcntr}
\counterwithin{equation}{section}
\fi

\ifC
\let\paragraph\subparagraph
\fi

\newcommand{\eps}{\varepsilon}
\newcommand{\adist}{\widetilde{\delta}}

\newcommand{\codestyle}[1]{\texttt{#1}}

\newcommand{\Query}{\mbox{\codestyle{Query}}}

\newcommand{\ecc}{\codestyle{ecc}}

\newcommand{\FarthestVertex}{\codestyle{Farthest-Vertex}}

\newcommand{\OrthRange}{\mathcal{OR}}

\newcommand{\Ot}{\tilde{O}}

\newcommand{\RR}{\mathbb{R}}

\newcommand{\ssin}[1]{P_{#1}^{\mathrm{in}}}
\newcommand{\ssout}[1]{P_{#1}^{\mathrm{out}}}

\newcommand{\Ssin}[1]{S_{#1}^{\mathrm{in}}}
\newcommand{\Ssout}[1]{S_{#1}^{\mathrm{out}}}

\newcommand{\psii}[2]{\psi^{#1}_{#2}}

\DeclarePairedDelimiter{\ceil}{\lceil}{\rceil}

\DeclarePairedDelimiter{\pair}{\langle}{\rangle}

\begin{document}
\ifC\else\pagenumbering{alph}\fi
\ActivateWarningFilters[pdftoc]
\maketitle

\ifC\else\thispagestyle{empty}\fi

\begin{abstract}
Computing the diameter of a graph is a fundamental problem in algorithmic graph theory. For weighted planar graphs, both directed and undirected, 
the fastest known exact algorithm, by Zheng and Prigan~[arXiv'26], runs in $\Ot(n^{3/2})$ time. 
While near-linear-time $(1+\eps)$-approximations have been known for undirected planar graphs for over a decade, starting with Weimann and Yuster~[ICALP'13, TALG'15], no such algorithm was known for directed planar graphs.

We present the first near-linear-time algorithm that approximates the diameter, all two-sided eccentricities, and the radius of directed planar graphs with nonnegative edge weights within a factor of $1+\eps$. The algorithm is deterministic and runs in $O(n\log^{O(1/\eps)}n)$ time.

\end{abstract}

\ifC\else
\clearpage
\pagenumbering{arabic}
\fi

\newpage
\section{Introduction}

The problem of computing (or approximating) the \emph{diameter} of a graph, that is, the maximum distance between any pair of vertices, is a central problem in graph algorithms (see, for example,~\cite{aingworth1999fast, backurs2021toward, DBLP:conf/soda/ChechikLRSTW14, dalirrooyfard2025hardness, dalirrooyfard2019tight, roditty2013fast}).

In general weighted graphs, the fastest algorithms for computing the diameter exactly  are via \emph{All-Pairs Shortest Paths} (APSP), and obtaining a faster algorithm is a major open problem in algorithmic graph theory.

Furthermore, Roditty and Vassilevska Williams~\cite{roditty2013fast} showed that under the Orthogonal Vectors Hypothesis (OVH)\footnote{The Orthogonal Vectors Hypothesis (OVH) states that for every $\gamma>0$ there is a constant $c$ such that, given two sets of $n$ vectors in $\{0,1\}^{c\log n}$, deciding whether there is an orthogonal pair with one vector from each set cannot be done in $O(n^{2-\gamma})$ time.}, for sparse graphs, there is no truly subquadratic time $(3/2-\gamma)$-diameter approximation algorithm, for any $\gamma > 0$.
 In particular, this rules out $(1+\eps)$-approximation algorithms with $\eps<1/2$ running in truly subquadratic time under OVH.

These barriers motivate the study of restricted graph classes. Planar graphs are among the most studied of these classes, in both exact and approximate settings. We next review the relevant results in planar graphs.

Frederickson~\cite{frederickson1987fast} gave the first optimal $O(n^2)$-time algorithm for APSP in directed weighted planar graphs, and thereby computed the diameter in the same time.
Later, Wulff-Nilsen~\cite{wulff2008wiener} provided the first indication that the diameter problem may be easier than the APSP problem in planar graphs, by presenting an $O(n^2 \frac{(\log\log{n})^4}{\log{n}})$-time algorithm for the problem. 
In a seminal paper, Cabello~\cite{cabello2018subquadratic} achieved truly subquadratic time, $\Ot(n^{11/6})$ expected\footnote{$\Ot(\cdot)$ omits polylogarithmic factors.}, using the heavy machinery of \emph{additively weighted Voronoi diagrams} for planar graphs.
Gawrychowski, Kaplan, Mozes, Sharir, and Weimann~\cite{gawrychowski2021voronoi} improved the running time to $\Ot(n^{5/3})$ deterministically, and very recently, Zheng and Prigan~\cite{zheng2026faster} improved it to $\Ot(n^{3/2})$. All these algorithms work for both directed and undirected weighted planar graphs.

\medskip
Despite this extensive line of work, exact algorithms for weighted planar graphs are still far from near-linear time.
A natural approach to achieving a near-linear running time is to settle for an approximation of the diameter. Weimann and Yuster~\cite{weimann2015approximating} were the first to show that the conditional lower bound for general graphs does not hold in the planar setting. 
They presented a $(1+\eps)$-diameter approximation algorithm that runs in $O(n\eps^{-4}\log^4 n+n2^{O(1/\eps)})$ time for weighted \textit{undirected} planar graphs.
The runtime was subsequently improved to $\Ot(n\,\mathrm{poly}(1/\eps))$ by Chan and Skrepetos~\cite{chan2019faster}, using the machinery of Voronoi diagrams for planar graphs, and was later also obtained via a different approach by Li and Parter~\cite{li2019planar}. These are the state-of-the-art algorithms for diameter approximation in weighted undirected planar graphs.
\begin{table}[t]
\centering
\begin{tabular}{llll}
\toprule
Graphs & Approximation & Running time & Reference \\
\midrule
\multirow{3}{*}{Directed} & \multirow{3}{*}{Exact}
    & $\Ot(n^{11/6})$ expected & \cite{cabello2018subquadratic} \\
  & & $\Ot(n^{5/3})$ & \cite{gawrychowski2021voronoi} \\
  & & $\Ot(n^{3/2})$ & \cite{zheng2026faster} \\
\midrule
\multirow{2}{*}{Undirected} & \multirow{2}{*}{$1+\eps$}
    & $O(n\eps^{-4}\log^4 n+n2^{O(1/\eps)})$ & \cite{weimann2015approximating} \\
  & & $\Ot(n\,\mathrm{poly}(1/\eps))$ & \cite{chan2019faster,li2019planar} \\
\midrule
Directed & $1+\eps$ & $O(n\log^{O(1/\eps)}n)$ & \Cref{thm:diameter} \\
\bottomrule
\end{tabular}
\caption{Selected diameter algorithms for weighted planar graphs.}
\label{tab:diameter}
\end{table}

Prior to this work, no near-linear-time $(1+\eps)$-approximation was known for the diameter of weighted directed planar graphs (see \Cref{tab:diameter}), leading to the following question.

\begin{question}\label{q:1}
    Is there a near-linear-time algorithm that computes a $(1+\eps)$-approximation to the diameter of a weighted directed planar graph?
\end{question}

\paragraph{Our results.}
In this paper, we answer \Cref{q:1} question in the affirmative by presenting a $(1 + \eps)$-diameter approximation algorithm that runs in $\Ot_\eps(n)$ time, for any $\eps > 0$.\footnote{$\Ot_\eps(\cdot)$ omits polylogarithmic factors whose exponent may depend on $1/\eps$.}

\begin{theorem}[Diameter]\label{thm:diameter}\Copy{thm:diameter}{
    For every $\eps>0$, there is a deterministic $O(n\log^{O(1/\eps)} n)$-time $(1+\eps)$-approximation algorithm for the diameter of directed planar graphs with $n$ vertices and nonnegative edge weights.}
\end{theorem}

Beyond the diameter, two fundamental and closely related problems that have received extensive attention are computing the two-sided eccentricity of every vertex $u\in V$ (the maximum of its incoming and outgoing distances), and the corresponding radius (the minimum two-sided eccentricity)~\cite{abboud2016approximation, backurs2021toward, chan2025diameter}. 
Our algorithmic approach naturally extends to these measures, yielding near-linear-time $(1+\eps)$-approximations for both.

\begin{theorem}[Eccentricities]\label{thm:eccentricities}\Copy{thm:eccentricities}{
    For every $\eps>0$, there is a deterministic $O(n \log^{O(1/\eps)} n)$-time algorithm that, given a directed planar graph $G$ with $n$ vertices and nonnegative edge weights, returns for every vertex $u$ an estimate $E(u)$ with $\ecc(u)\le E(u)\le(1+\eps)\ecc(u)$, where $\ecc(u)$ is the two-sided eccentricity of $u$ in $G$.}
\end{theorem}

Since the diameter and the radius are the maximum and the minimum eccentricity, \Cref{thm:eccentricities} implies \Cref{thm:diameter} and the following corollary.

\begin{corollary}[Radius]\label{cor:radius}
    For every $\eps>0$, there is a deterministic $O(n \log^{O(1/\eps)} n)$-time $(1+\eps)$-approximation algorithm for the two-sided radius of directed planar graphs with $n$ vertices and nonnegative edge weights.
\end{corollary}

\paragraph{Techniques.}
Roughly speaking, all the near-linear undirected approximation algorithms of~\cite{chan2019faster,li2019planar,weimann2015approximating} consider paths that go through a single, globally shared set of $O(1/\eps)$ portals (special vertices) on a shortest-path separator. A shortest-path separator consists of $O(1)$ shortest paths whose removal splits the graph into balanced parts, so every path between two parts meets one of these paths.

Two difficulties arise in directed graphs (digraphs). The first concerns the separators. To compute the largest distance from a source $u$, we assign $u$ and its destinations to separators, and we need every shortest path between $u$ and an assigned destination, in either direction, to meet the separator. Otherwise, the route through the separator may be much longer than the actual distance, and a single such overestimate can spoil the maximum. The shortest-path separators of Thorup~\cite{thorup2004compact} for planar digraphs do not guarantee this: a shortest path may miss the separator of a subgraph that contains both of its endpoints. A distance oracle tolerates this, since it takes the minimum over several separators, but a maximum over destinations does not. We therefore develop a new \emph{in-out separator system}, in which every pair of vertices is assigned to some separator, the shortest paths between assigned pairs meet that separator, and every vertex is a source of only $O(\log n)$ separators (\Cref{lem:shortest-path-system}).

The second difficulty is that a single shared set of portals does not work in digraphs. In undirected graphs, a shortest path from $u$ to $v$ that meets the separator at a vertex $x$ can detour from $x$ to a nearby portal and back at a small extra cost. In digraphs, one of these two parts goes against the direction of the separator path, so this detour may not exist. Thorup's $(1+\eps)$-approximate distance oracle for planar digraphs~\cite{thorup2004compact} resolves this by giving every vertex its own $O(1/\eps)$ \emph{in-portals}, where paths from it enter the separator, and $O(1/\eps)$ \emph{out-portals}, where paths to it leave the separator. The distance from $u$ to $v$ is then approximated by the shortest route that goes from $u$ to one of its in-portals, continues along the separator, and reaches $v$ from one of its out-portals. Hence the largest distance from $u$ becomes a maximum over destinations $v$ of a minimum over pairs of portals, one of $u$ and one of $v$. The data structures built for shared portals do not handle this $\max\min$ problem.

The length of a path through an in-portal of the source and an out-portal of the destination splits into a term that depends only on the source and a term that depends only on the destination. The order in which the portals of the two vertices interleave along the separator fixes which out-portal of the destination is paired with each in-portal of the source. Together with the split, this reduces the $\max\min$ optimization to orthogonal range queries in $O(1/\eps)$ dimensions: every vertex is embedded as a point in $\RR^{O(1/\eps)}$, and for every interleaving and every in-portal, the destinations for which that in-portal is optimal correspond to an axis-parallel box. Since the number of interleavings depends only on $\eps$, the queries take $\Ot_\eps(1)$ time per vertex, which gives a near-linear-time algorithm for the diameter.

Interestingly, the idea of using orthogonal range searching for $\max\min$ optimizations in the context of diameter computation is not new: it dates back to Cabello and Knauer~\cite{cabello2009treewidth}, and was later used by Abboud, Vassilevska Williams, and Wang~\cite{abboud2016approximation}, in both cases for computing the diameter \emph{exactly} in graphs of bounded treewidth. Shifting to the \emph{approximation} setting allows us to harness this classical technique for planar graphs, whose treewidth may be unbounded.

\paragraph{Related work.}
Exact diameter algorithms are also known for unweighted undirected planar graphs~\cite{abboud2023voronoi,fredslund2026tight}. Near-linear $(1+\eps)$-approximations of the diameter have been extended beyond planar graphs to graphs excluding a fixed apex minor~\cite{wlodarczyk2025beyond}, and exact truly subquadratic algorithms are known for real-weighted minor-free digraphs~\cite{zheng2026realweighted}. Voronoi diagrams for planar graphs have also led to compact distance oracles~\cite{charalampopoulos2023almost, charalampopoulos2022exact, cohen2017fast, gawrychowski2018better}.

\paragraph{Organization.}
\Cref{sec:prelims} gives the preliminaries, including our new in-out separator system. \Cref{sec:overview} gives an overview of the algorithm. \Cref{sec:data-structure} presents the data structure, and \Cref{sec:algo} presents the algorithm. \Cref{sec:in-out-separators,sec:portal-preprocessing} prove the separator system and the portal construction.

\section{Preliminaries}\label{sec:prelims}
Let $G = (V, E, w)$ be a directed planar graph with $n$ vertices, nonnegative real edge weights, and no parallel arcs, so $|E|=O(n)$. We assume that $G$ is strongly connected, since otherwise all eccentricities are infinite, which is detected in linear time. We use unit-cost arithmetic, comparisons, and rounding of scaled weights. We assume $0<\eps\le 1$. A $(1+\eps)$-approximation of a quantity $D$ is an estimate in $[D,(1+\eps)D]$. 
For two vertices $u, v \in V$, we use $P(u,v)$ to denote a \emph{shortest path} from $u$ to $v$, and $\delta(u,v)$ to denote
the \emph{length} of $P(u, v)$, i.e., \(\delta(u,v)=\sum_{e\in P(u,v)} w(e)\). 
For a shortest path $P(u,v)$ and two vertices $x,y\in P(u,v)$ we denote by $\delta_P(x, y)$ the distance from $x$ to $y$ along $P$. If $x$ appears after $y$ in $P(u,v)$ then $\delta_P(x,y)=\infty$.

For a path $P$, we let $P[i]$ denote the $i$-th vertex in $P$ (starting with $i = 1$), and we say that $i$ is the \emph{index} of the vertex $P[i]$ in $P$. Let $|P|$ be the number of vertices in $P$, and let $[k]$ be the set $\{1,2,3,\dots,k\}$. 

Let $u\in V$ be a vertex. The \emph{eccentricity} of $u$, $\ecc(u)$, is defined as the length of the
longest shortest path that has $u$ as an endpoint. That is,
\(\ecc(u) = \max_{v\in V}\{\delta(u,v), \delta(v,u)\}\). The \emph{diameter} of $G$, $\Delta(G)$, is defined to be the maximum distance between two vertices in $G$, or, equivalently,
as the maximum eccentricity of a vertex in $G$, i.e., \(\Delta(G) = \max_{u,v \in V} \delta(u,v) = \max_{u \in V} \ecc(u)\). The radius of $G$ is defined as $\min_{u\in V}\ecc(u)$. When $G$ is clear from context, it is omitted.

The following simple lemma gives connections between $\ecc(u)$ and the diameter.
\begin{lemma} \label{lem:ecc-diameter}
    Let $G = (V, E, w)$ be a weighted directed graph whose diameter is $\Delta$. For every $u\in V$,
    $\Delta \leq 2\cdot \ecc(u) \leq 2\Delta$.
\end{lemma}
\begin{proof}
    By definition, $\ecc(u)\le\Delta$. Let $a,b\in V$ satisfy $\delta(a,b)=\Delta$. By the triangle inequality, $\Delta=\delta(a,b)\le\delta(a,u)+\delta(u,b)\le2\ecc(u)$.
\end{proof}

From \Cref{lem:ecc-diameter}, the following folklore algorithm provides an efficient, simple way to approximate the diameter within a factor of $2$. Throughout the paper, we denote by $\hat{\Delta}$ a value with $\hat\Delta\le\Delta\le2\hat\Delta$. 
\begin{lemma}\label{lem:2-approx}
    Given a directed planar graph $G$ with $n$ vertices, we can compute in $O(n)$ time a value $\hat{\Delta}$ with $\hat\Delta\le\Delta\le2\hat\Delta$, where $\Delta=\Delta(G)$.
\end{lemma}
\begin{proof}
Let $u\in V$ be an arbitrary vertex. The algorithm computes the distances in $G$ from and to $u$, and returns their maximum $\hat\Delta=\ecc(u)$. This takes $O(n)$ time, using the single-source shortest path algorithm of Henzinger, Klein, Rao, and Subramanian~\cite{henzinger1997faster}.
From~\Cref{lem:ecc-diameter}, $\frac{1}{2}\Delta \leq \ecc(u) \leq \Delta$.
\end{proof}

We use the following standard range tree, in which every canonical subset stores the maximum weight of its points.
\begin{lemma}[Range tree for range maxima, cf.\ {\cite[Theorem~5.9]{de2008computational}}] \label{lem:range-tree}
    Let $Q$ be a set of $n$ weighted points in $\RR^d$, where $d\geq 2$. There is a data structure that, given an axis-parallel box, returns the maximum weight of a point of $Q$ in the box in $2^{O(d)}\log^d n$ time. It uses $2^{O(d)}n\log^{d-1} n$ space and can be computed in $2^{O(d)}n\log^{d-1} n$ time.
\end{lemma}

For a shortest path $P$ and a vertex $u$, the \emph{in-portals} of $u$ on $P$ are vertices through which paths from $u$ enter $P$, and the \emph{out-portals} of $u$ are vertices through which paths to $u$ leave $P$.

\begin{lemma}[Portals~{\cite[Lemma~3.13]{thorup2004compact}}] \label{lem:portal}
    Let $\hat\Delta$ satisfy $\frac12\Delta\le\hat\Delta\le\Delta$, and let $P$ be a shortest path in $G$.
    For every vertex $u\in V$, there are sets $\ssin{u},\ssout{u}\subseteq P$ of $O(1/\eps)$ vertices each with the following property. For every $x\in P$, some $s\in \ssin{u}$ and $s'\in \ssout{u}$ satisfy:
    \begin{itemize}
        \item $\delta(u, s) + \delta_P(s, x) \leq \delta(u, x) + \eps \hat \Delta$
        \item $\delta_P(x, s') + \delta(s', u) \leq \delta(x, u) + \eps \hat \Delta$.
    \end{itemize}
\end{lemma}

We assume that $\ssin{u}$ and $\ssout{u}$ are sorted along $P$ and contain the first and last vertices of $P$, and we denote by $\ssin{u}[i]$ and $\ssout{u}[i]$ their $i$-th entries. In \Cref{sec:portal-preprocessing} we compute these lists for all the paths of our separator system (\Cref{lem:global-portal-construction}).

\begin{figure}[t]
    \centering
    \begin{tikzpicture}[
        >=stealth,
        vtx/.style={circle,draw=black,fill=black,inner sep=1.4pt},
        svtx/.style={circle,draw=red!75!black,fill=red!75!black,inner sep=1pt},
        plabel/.style={font=\scriptsize,inner sep=1pt}
    ]
        \fill[blue!8] (0,0) .. controls (-0.8,1) and (-1.2,2) .. (-1.1,3.2)
            -- (1.1,3.2) .. controls (1.2,2) and (0.8,1) .. (0,0);
        \fill[orange!12] (0,0) .. controls (-0.8,1) and (-1.2,2) .. (-1.1,3.2)
            -- (-3,3.2) -- (-3,0) -- cycle;
        \fill[orange!12] (0,0) .. controls (0.8,1) and (1.2,2) .. (1.1,3.2)
            -- (3,3.2) -- (3,0) -- cycle;
        \draw[gray!60] (-3,0) rectangle (3,3.2);

        \draw[thick,red!75!black,->]
            (0,0) .. controls (-0.8,1) and (-1.2,2) .. (-1.1,3.2);
        \draw[thick,red!75!black,->]
            (0,0) .. controls (0.8,1) and (1.2,2) .. (1.1,3.2);
        \foreach \x/\y in {0/0,0.29/0.39,0.56/0.81,0.76/1.21,0.94/1.68,1.06/2.17,1.11/2.67} {
            \node[svtx] at (\x,\y) {};
            \node[svtx] at (-\x,\y) {};
        }

        \node[blue!70!black] at (0,1.0) {$I_S$};
        \node[orange!80!black] at (-2.05,1.15) {$O_S$};
        \node[orange!80!black] at (2.1,1.15) {$O_S$};
        \node[red!75!black] at (-1.1,0.95) {$S$};
        \foreach \p in {(-0.7,2.75),(0.3,0.55),(-0.45,1.3),(0.55,1.35)}
            \fill[blue!70!black] \p circle (1.4pt);
        \foreach \p in {(-2.5,2.6),(-1.6,2.6),(-2.6,0.45),(-1.5,0.45),(-2.55,1.85),
                        (1.45,0.4),(2.65,0.45),(2.7,1.65),(1.55,2.95),(2.7,2.95)}
            \fill[orange!80!black] \p circle (1.4pt);

        \node[vtx,label=left:$u$] (u) at (-0.3,1.85) {};
        \node[vtx,label=right:$v$] (v) at (2.3,2.35) {};
        \draw[->,semithick] (u) to[out=-10,in=185] (0.94,1.68)
            to[out=5,in=220] node[pos=0.55,below right,plabel] {$P(u,v)$} (v);
        \draw[->,semithick] (v) to[out=160,in=-5]
            node[pos=0.3,above,plabel] {$P(v,u)$} (1.11,2.67)
            to[out=175,in=70] (u);
    \end{tikzpicture}
    \caption{A separator set $S$ (red) with its in-set $I_S$ and out-set $O_S$.}
    \label{fig:in-out-example}
\end{figure}
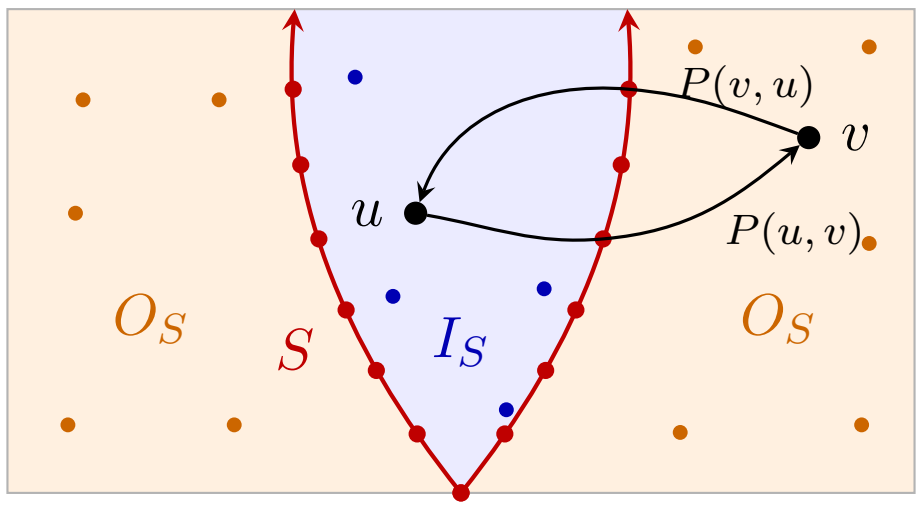
\subsection{Shortest-Path Separator Systems}\label{sec:separator-systems}

Thorup~\cite{thorup2004compact} introduced a \emph{shortest-path separator} for directed planar graphs, which has since become a central tool in algorithms for directed planar graphs (see, for example,~\cite{charalampopoulos2022single, cohen2017fast, le2022optimal}).
We use a different separator system in which each separator set $S$ of $O(1)$ shortest paths is associated with an \emph{in-set} $I_S$ and an \emph{out-set} $O_S$, such that for every $u\in I_S$ and $v\in O_S$, both $P(u,v)$ and $P(v,u)$ intersect $S$ (\Cref{fig:in-out-example}). The algorithm uses the vertices of $I_S$ as sources and the vertices of $O_S$ as their destinations.
Unlike the separator system of Thorup~\cite{thorup2004compact}, our construction does not recurse on subgraphs. It keeps $G$ and a single shortest-path tree fixed and recurses on sets of vertices, so all separator paths are root-to-vertex paths of the same tree.

\begin{lemma}[The in-out shortest-path separator system]

\label{lem:shortest-path-system}\Copy{lem:shortest-path-system}{
Let $G=(V,E,w)$ be a strongly connected directed planar graph with $n$
vertices and nonnegative edge weights.  There is an indexed collection
$\mathcal S$ of \emph{separator sets} with the following properties.
Each set $S\in\mathcal S$ is a set of at most three directed
shortest paths in $G$.
Let $V(S)=\bigcup_{P\in S}V(P)$.

Associated with every set $S\in\mathcal S$ are sets
$I_S,O_S\subseteq V$, called
the \emph{in-set} and the \emph{out-set} of $S$. 
For every $u\in V$, let $\mathcal S(u)=\{S\in\mathcal S \mid u\in I_S\}$.
The following properties hold:
\begin{enumerate}
    \item For every two distinct vertices $u,v\in V$ there is an $S \in \mathcal S$ 
    such that $u\in I_S$ and $v\in O_S$.
    \item For every $S\in\mathcal S$, every $u\in I_S$, and every
    $v\in O_S$, it holds that
    $P(u,v)\cap V(S)\neq\emptyset$ and
    $P(v,u)\cap V(S)\neq\emptyset$.
    \item For every $u\in V$, $|\mathcal S(u)|=O(\log n)$, and
    $\sum_{S\in\mathcal S}(|I_S|+|O_S|)=O(n\log n)$.
\end{enumerate}
Moreover, $|\mathcal S|=O(n)$, all the paths of all the sets in $\mathcal S$ are root-to-vertex paths of a single shortest-path tree $T$ of $G$, and in $O(n\log n)$ time one can compute $T$, $\mathcal S$, and all the
in-sets and out-sets.
}
\end{lemma}
The same set of paths may appear in $\mathcal S$ under several indices, with different in-sets and out-sets.
The construction recursively splits a set of active vertices, starting with $V$, by three root paths of $T$ into sides that contain at most half of the active vertices each. A pair $u,v$ is covered at the deepest recursive step at which both are active. At this step either $u$ or $v$ lies on the paths, or they lie in different sides, so every path between them meets one of the paths.
Due to space limitations, we defer the proof of \Cref{lem:shortest-path-system} to \Cref{sec:in-out-separators}.

\section{Technical Overview}\label{sec:overview}

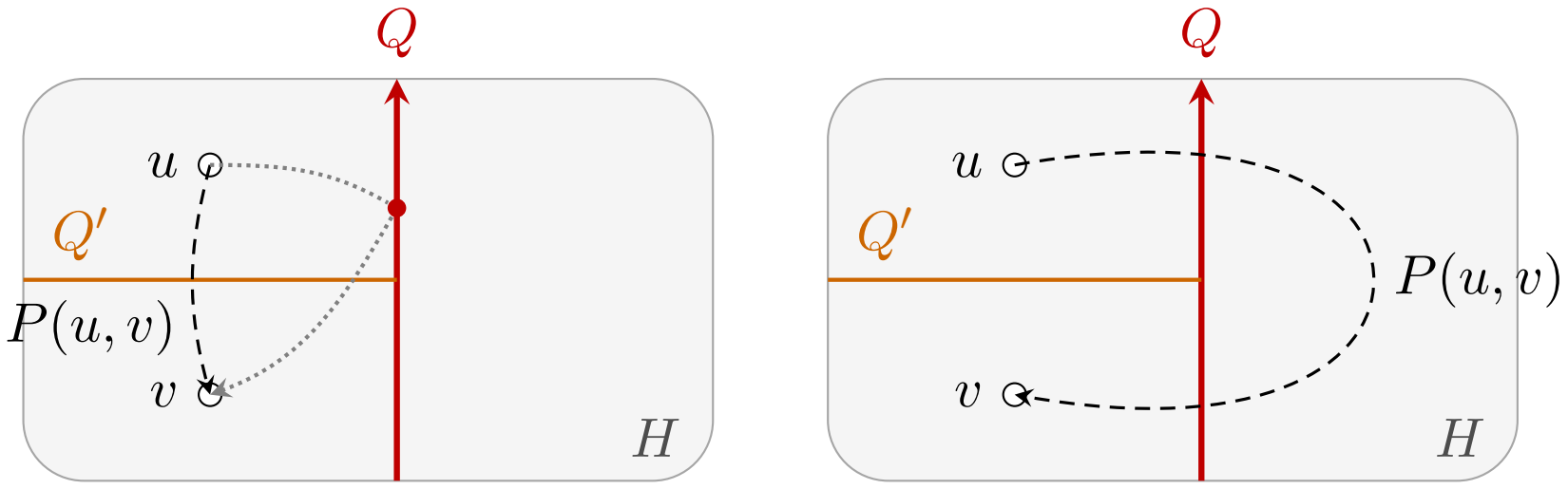
\begin{figure}[t]
    \centering
    \begin{tikzpicture}[
        >=stealth,
        vertex/.style={circle,draw=black,fill=white,inner sep=1.6pt}
    ]
        \foreach \dx/\panel in {0/(a), 5.6/(b)} {
        \begin{scope}[xshift=\dx cm]
            \draw[rounded corners=12pt,draw=gray!70,fill=gray!8] (-2.6,-1.4) rectangle (2.2,1.4);
            \node[gray!60!black] at (1.8,-1.1) {$H$};
            \draw[very thick,red!75!black,->] (0,-1.4) -- (0,1.4) node[above] {$Q$};

            \draw[thick,orange!80!black] (-2.6,0) -- (0,0);
            \node[orange!80!black,above] at (-2.2,0) {$Q'$};

            \node[vertex,label=left:$u$] (u) at (-1.3,0.8) {};
            \node[vertex,label=left:$v$] (v) at (-1.3,-0.8) {};
            \node at (-0.2,-1.9) {\panel};
        \end{scope}
        }

        \node[circle,fill=red!75!black,inner sep=1.3pt] (x) at (0,0.5) {};
        \draw[->,densely dashed,semithick] (-1.3,0.8) to[bend right=15]
            node[pos=0.72,left] {$P(u,v)$} (-1.3,-0.8);
        \draw[->,densely dotted,thick,gray] (-1.3,0.8) to[out=0,in=150] (x)
            to[out=-120,in=20] (-1.3,-0.8);

        \begin{scope}[xshift=5.6cm]
            \draw[->,densely dashed,semithick] (-1.3,0.8)
                .. controls (0.4,1.1) and (1.2,0.6) .. (1.2,0)
                node[right] {$P(u,v)$}
                .. controls (1.2,-0.6) and (0.4,-1.1) .. (-1.3,-0.8);
        \end{scope}
    \end{tikzpicture}
    \caption{Vertices $u,v$ on the same side of the separator $Q$ of a subgraph $H$ of Thorup's recursion. (a) $P(u,v)$ misses $Q$ and meets the deeper separator $Q'$, and a route through $Q$ (dotted) may be much longer. (b) $P(u,v)$ meets $Q$.}
    \label{fig:thorup-miss}
\end{figure}

Our goal is to compute, for every vertex $u$, the value $\max_{v\in V}\delta(u,v)$ up to an additive error of $O(\eps\Delta)$, which suffices since $\ecc(u)\ge\Delta/2$ (\Cref{lem:ecc-diameter}). Distances into $u$ are handled in the same way in the reverse graph.
For simplicity, in this overview we assume that $\Delta$ is known (the algorithm uses $\hat\Delta=\ecc(w)$, for some $w\in V$ instead), and we fix sets $I,O\subseteq V$ such that every shortest path from a vertex of $I$ to a vertex of $O$ meets a single shortest path $P$. The full algorithm takes such sets from the in-out separator system of \Cref{lem:shortest-path-system}: an in-set $I_S$ and an out-set $O_S$ of a separator set $S$ of at most three paths (\Cref{fig:in-out-example}). With several paths, the portal lists below are concatenated over the paths of $S$. Every vertex lies in $O(\log n)$ in-sets.
The shortest-path decomposition of Thorup~\cite{thorup2004compact} does not suffice here. In this decomposition, $P(u,v)$ meets the separator of some subgraph of the recursion, namely one in which $u$ and $v$ are separated. In every other subgraph that contains both $u$ and $v$, the two vertices lie on the same side of the separator, so $P(u,v)$ may or may not meet it, and the algorithm cannot tell which (\Cref{fig:thorup-miss}). A route through a separator that $P(u,v)$ misses may be much longer than $\delta(u,v)$. A distance oracle minimizes over these separators, so such an overestimate is harmless if another separator supplies an accurate estimate. Our algorithm instead maximizes over all the destinations in an out-set. An overestimate for even one destination can invalidate that maximum. We therefore need separator sets $S$ such that $P(u,v)$ meets $S$ for every $u\in I_S$ and $v\in O_S$, so that the additive upper bound holds for every such pair.

\begin{figure}[t]
    \centering
    \tikzset{every picture/.style={line width=0.75pt}} 
    \begin{adjustbox}{max width=\linewidth}
    \input{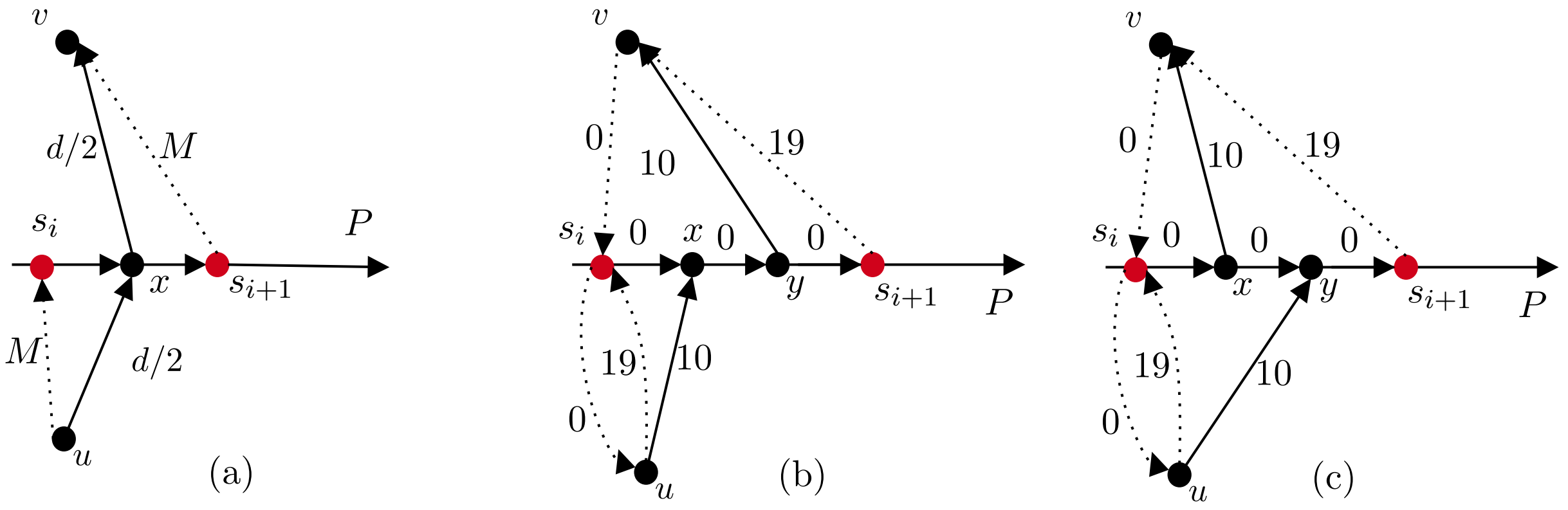}
    \end{adjustbox}

    \caption{Shared portals $s_i,s_{i+1}$. (a) Arcs of $P$ have weight $0$. A route through a portal costs $M+d/2$, although $\delta(u,v)=d$. (b),(c) The same distances $\delta(u,s)$ and $\delta(s,v)$ for every portal $s$, but $\delta(u,v)$ is $20$ and $29$.}

    \label{fig:problem-directed}
\end{figure}

\paragraph{Previous undirected approximations.}
Previous algorithms for weighted undirected planar graphs~\cite{chan2019faster,li2019planar,weimann2015approximating} choose
a set $C\subseteq P$ of $O(1/\eps)$ vertices such that, for every $u\in I$ and $v\in O$,
\[
\delta(u,v) \le \min_{s\in C}(\delta(u,s)+\delta(s,v)) \le \delta(u,v) + \eps\Delta.
\]
For example, it suffices that every vertex of $P$ is within distance $\eps\Delta/2$ of $C$. If $P(u,v)$ meets $P$ at $x$ and $s\in C$ is closest to $x$, the route $u\to x\to s\to x\to v$ is at most $\eps\Delta$ longer than $\delta(u,v)$.
Because $C$ is shared by all vertices, these algorithms build data structures that answer $\max_{v\in O}\min_{s\in C}(\delta(u,s)+\delta(s,v))$ in $\Ot_\eps(1)$ time for a given $u$.

\paragraph{Where does this approach fail for directed planar graphs?}
In directed graphs, the inequality
$\min_{s\in C}\bigl(\delta(u,s)+\delta(s,v)\bigr) \le \delta(u,v) + \eps\Delta$
does \emph{not} hold in general.
Consider a shortest path from $u$ to $v$ of length $d$ that meets $P$ only at a vertex $x$, and let $s_i,s_{i+1}\in C$ be the portals before and after $x$.
Then routing through $s_i$ or $s_{i+1}$ may be arbitrarily more expensive than $d$ (see \Cref{fig:problem-directed}(a)). The undirected detour from $x$ to a portal and back fails, since one of its two parts goes against the direction of $P$.
No choice of $C$ avoids this. For example, consider a directed cycle in which one arc has weight $\Delta$ and all other arcs have weight zero, and let $P$ be the path formed by the zero-weight arcs. Every set $C$ of fewer than a third of the vertices misses two consecutive vertices $x,y$ of $P$. Then $\delta(x,y)=0$, whereas every route from $x$ to $y$ through a vertex of $C$ uses the heavy arc and has length $\Delta$.
Moreover, endpoint distances to fixed portals need not determine the pairwise distance (\Cref{fig:problem-directed}(b) and~(c)).

\begin{figure}[t]
    \centering
    \begin{tikzpicture}[
        >=stealth,
        inp/.style={circle,draw=blue!70!black,fill=blue!25,inner sep=1.8pt},
        outp/.style={rectangle,draw=orange!80!black,fill=orange!35,inner sep=2.2pt},
        vertex/.style={circle,draw=black,fill=white,inner sep=1.6pt}
    ]
        \draw[thick,->] (0,-2) -- (0,2.2) node[above] {$P$};

        \node[vertex,label=right:$u$] (u) at (2.4,0.1) {};
        \node[vertex,label=left:$v$] (v) at (-2.4,0.5) {};
        \node[circle,fill=black,inner sep=1.3pt] (x) at (0,0.2) {};
        \node[below right=-1pt] at (x) {$x$};
        \draw[->,densely dashed,gray] (u) to[out=175,in=-5] (x)
            to[out=175,in=-10] node[pos=0.5,below] {$P(u,v)$} (v);

        \draw[line width=2pt] (0,-0.8) -- (0,1.1);
        \draw[->,thick,blue!70!black] (u) to[out=-150,in=-20] (0.1,-0.83);
        \draw[->,thick,orange!80!black] (0,1.1) to[out=160,in=30] (v);

        \foreach \y in {-1.6,0.7,1.6} \node[inp] at (0,\y) {};
        \node[inp,label=left:$s$] at (0,-0.8) {};
        \foreach \y in {-1.2,-0.3,1.9} \node[outp] at (0,\y) {};
        \node[outp,label=right:$t$] at (0,1.1) {};
    \end{tikzpicture}
    \caption{In-portals of $u$ (circles) and out-portals of $v$ (squares). The route $u\to s\to t\to v$ approximates $\delta(u,v)$.}
    \label{fig:portals}
\end{figure}
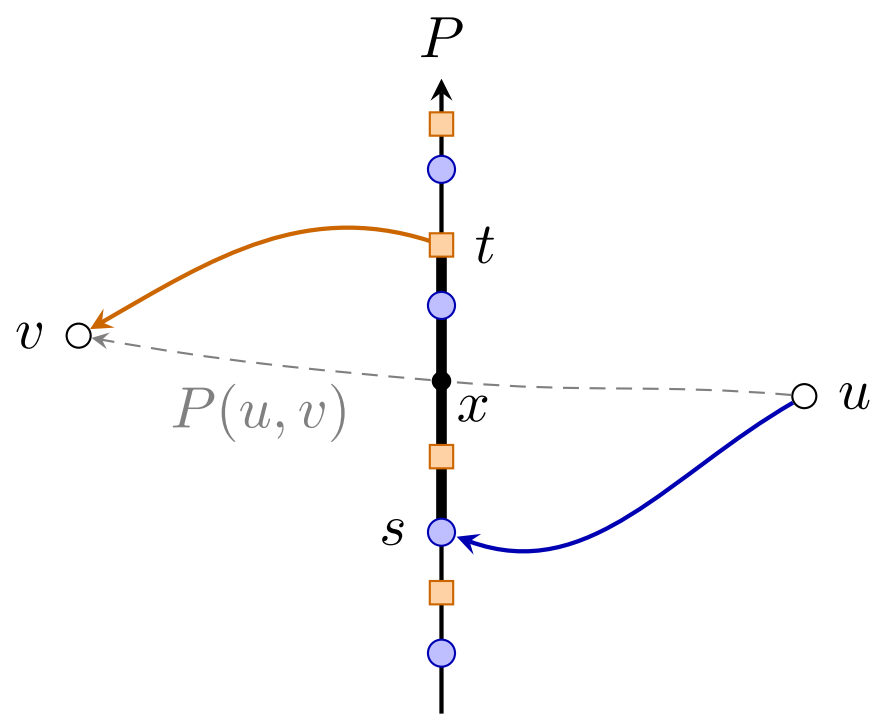

\paragraph{Thorup's portals.}
Thorup~\cite{thorup2004compact} gives every vertex $u$ its own in-portals $\ssin{u}$ and out-portals $\ssout{u}$ on $P$, and shows that for every $u\in I$ and $v\in O$ (\Cref{fig:portals}):
\[
\delta(u,v) \le \min_{\pair{s,t}\in \ssin{u}\times \ssout{v}}(\delta(u,s)+\delta_P(s,t)+\delta(t,v)) \le \delta(u,v) + 2\eps\Delta,
\]
where $\delta_P(s,t)=\infty$ unless $t$ is at or after $s$ on $P$. The upper bound follows by applying \Cref{lem:portal} to a vertex $x\in P(u,v)\cap P$, which gives an in-portal $s$ of $u$ at or before $x$ and an out-portal $t$ of $v$ at or after $x$.
The resulting $\max\min$ problem, $\max_{v\in O}$ of this expression, ranges over pairs of vertex-dependent portal sets, and the shared-portal data structures no longer apply.

\begin{figure}[t]
    \centering
    \begin{adjustbox}{max width=\linewidth}
    \begin{tikzpicture}[
        >=stealth,
        inp/.style={circle,draw=blue!70!black,fill=blue!25,inner sep=1.8pt},
        outp/.style={rectangle,draw=orange!80!black,fill=orange!35,inner sep=2.2pt},
        vertex/.style={circle,draw=black,fill=white,inner sep=1.6pt}
    ]
        \draw[thick,->] (0,0) -- (6.9,0) node[right] {$P$};
        \node[circle,fill=black,inner sep=1.2pt,label=left:$r$] at (0,0) {};

        \node[inp] at (0.8,0) {};
        \node[inp,label={below left:$s$}] (s) at (2.0,0) {};
        \node[inp] at (4.7,0) {};
        \node[outp] at (1.4,0) {};
        \node[outp,label={above left:$N(s,v)$}] (N) at (4.0,0) {};
        \node[outp,label={above right:$t$}] (t) at (5.7,0) {};

        \node[vertex,label=above:$u$] (u) at (1.1,1.6) {};
        \node[vertex,label=above:$v$] (v) at (5.0,1.8) {};

        \draw[->,thick,blue!70!black] (u) to[out=-60,in=110]
            node[pos=0.45,left] {$\delta(u,s)$} (s);
        \draw[line width=2pt] (s) -- (N);
        \draw[->,thick,orange!80!black] (N) to[out=70,in=-130]
            node[pos=0.6,left] {$\delta(N(s,v),v)$} (v);
        \draw[->,densely dashed,gray!70] (t) to[out=110,in=-40] (v);

        \draw[decorate,decoration={brace,mirror,amplitude=5pt},blue!70!black]
            (0,-0.45) -- (2.0,-0.45) node[midway,below=6pt] {$L(s)$};
        \draw[decorate,decoration={brace,mirror,amplitude=5pt},orange!80!black]
            (0,-1.15) -- (4.0,-1.15) node[midway,below=6pt] {$L(N(s,v))$};

        \node at (3.45,-2.15) {(a)};

        \begin{scope}[xshift=8.4cm]
            \foreach \y/\name/\oa/\ob/\oc/\od in
                {1.25/v/1.3/4.6/5.2/7.1, -0.25/v'/2.0/4.2/5.5/6.9} {
                \draw[thick,->] (0,\y) -- (7.5,\y);
                \node[left] at (-0.1,\y) {$\name$};
                \node[inp,label={[inner sep=1pt]below:$\iota_1$}] (i1) at (0.5,\y) {};
                \node[inp,label={[inner sep=1pt]below:$\iota_2$}] (i2) at (2.7,\y) {};
                \node[inp,label={[inner sep=1pt]below:$\iota_3$}] (i3) at (3.6,\y) {};
                \node[inp,label={[inner sep=1pt]below:$\iota_4$}] (i4) at (6.2,\y) {};
                \node[outp,label={[inner sep=1pt]below:$o_1$}] (o1) at (\oa,\y) {};
                \node[outp,label={[inner sep=1pt]below:$o_2$}] (o2) at (\ob,\y) {};
                \node[outp,label={[inner sep=1pt]below:$o_3$}] (o3) at (\oc,\y) {};
                \node[outp,label={[inner sep=1pt]below:$o_4$}] (o4) at (\od,\y) {};
                \draw[->] (i1) to[bend left=45] (o1);
                \draw[->] (i2) to[bend left=40] (o2);
                \draw[->] (i3) to[bend left=50] (o2);
                \draw[->] (i4) to[bend left=50] (o4);
            }
            \node at (3.7,-1.05) {$p_\pi=(1,2,2,4)$};
            \node at (3.7,-2.15) {(b)};
        \end{scope}
    \end{tikzpicture}
    \end{adjustbox}
    \caption{Blue circles are in-portals of $u$ and orange squares are out-portals of the destination. (a) The route through the successor $N(s,v)$, a route through a later out-portal $t$, and the root distances $L$. (b) Two destinations with the same interleaving $\pi=\iota_1\,o_1\,\iota_2\,\iota_3\,o_2\,o_3\,\iota_4\,o_4$.}
    \label{fig:successor-interleaving}
\end{figure}
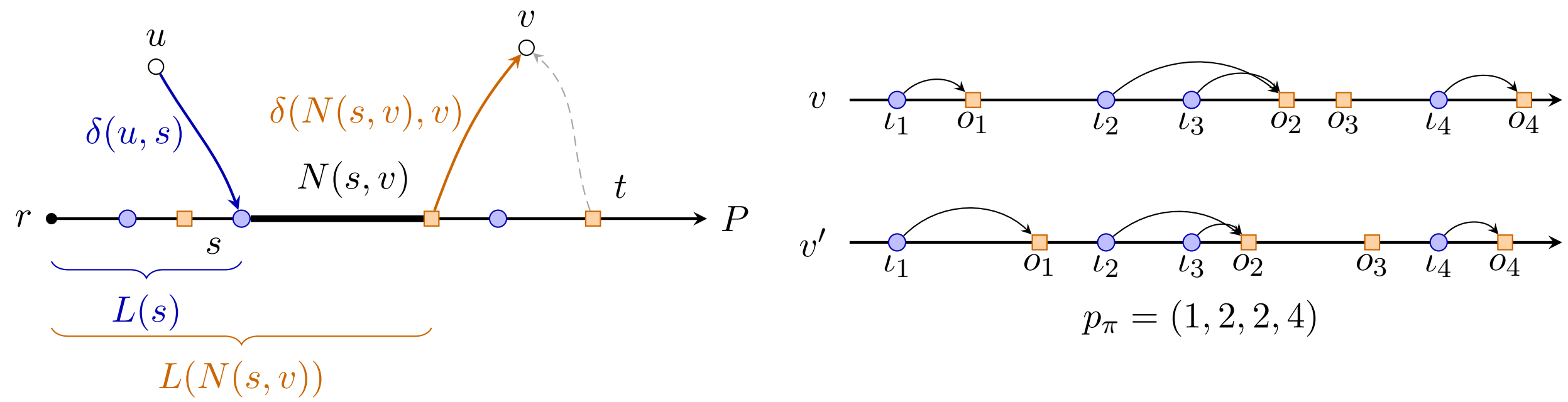

\paragraph{Our approach.}
We make three observations.
\begin{enumerate}
    \item \emph{Successors suffice.} For a fixed in-portal $s$, it suffices to consider the first out-portal $N(s,v)$ of $v$ at or after $s$ (\Cref{fig:successor-interleaving}(a)): by the triangle inequality, any later out-portal $t$ yields a route that is at least as long as the route through $N(s,v)$ (cf.\ \cite[Lemma~3.6]{thorup2004compact}).
    \item \emph{The separator term splits.} Let $r$ be the first vertex of $P$, and let $L(x)=\delta(r,x)$. For $t$ at or after $s$ on $P$ we have $\delta_P(s,t)=L(t)-L(s)$ (\Cref{fig:successor-interleaving}(a)), and therefore
    \[
        \delta(u,s)+\delta_P(s,t)+\delta(t,v)=\underbrace{\bigl(\delta(u,s)-L(s)\bigr)}_{a_u(s)}+\underbrace{\bigl(L(t)+\delta(t,v)\bigr)}_{b_v(t)}.
    \]
    The routed length is the sum of a term depending only on $u$ and a term depending only on $v$.
    \item \emph{Interleavings fix successor indices.} Let $k=O(1/\eps)$ bound the sizes of the portal lists. The index $p_\pi(i)$ of $N(\ssin{u}[i],v)$ in $\ssout{v}$ is determined by the interleaving pattern $\pi$ of $\ssin{u}$ and $\ssout{v}$, that is, the order of their entries along $P$, recording only which list each entry belongs to (\Cref{fig:successor-interleaving}(b)). There are only $2^{O(1/\eps)}$ such patterns.
\end{enumerate}
Fix $u$, a pattern $\pi$, and an index $i$, and consider the vertices $v$ that induce $\pi$ and for which $\ssin{u}[i]$ attains $\min_{i'}\bigl(a_u(\ssin{u}[i'])+b_v(\ssout{v}[p_\pi(i')])\bigr)$. The condition on $\pi$ is a set of inequalities between the positions of the out-portals of $v$ and the positions of the in-portals of $u$, and the minimality of $i$ is the set of inequalities
\[
    b_v(\ssout{v}[p_\pi(i)])-b_v(\ssout{v}[p_\pi(i')])\le a_u(\ssin{u}[i'])-a_u(\ssin{u}[i]),\qquad i'\in[k].
\]
In all these inequalities, one side depends only on $v$ and the other only on $u$. Hence, after embedding every $v$ as the point in $\RR^{2k}$ of its positions and of its label differences relative to the index $q=p_\pi(i)$, the points of these vertices are exactly the points in an axis-parallel box, and, since $a_u(\ssin{u}[i])$ is fixed, the maximum routed length over these vertices is a range-maximum query for $b_v(\ssout{v}[p_\pi(i)])$.

For example, let $k=2$ and $\pi=\iota_1\,o_1\,\iota_2\,o_2$, so $p_\pi=(1,2)$, and let $a_u(\ssin{u}[1])=5$ and $a_u(\ssin{u}[2])=2$. A destination $v$ that induces $\pi$, with labels $b_j=b_v(\ssout{v}[j])$, has routed length $\min(5+b_1,\,2+b_2)$. The first in-portal attains the minimum if $b_1-b_2\le-3$, and the second if $b_2-b_1\le3$. These are the label constraints of the two boxes. Two destinations with labels $(b_1,b_2)=(6,11)$ and $(9,10)$ lie in the first and in the second box, respectively. Hence the two queries return $5+6=11$ and $2+10=12$, and the farthest estimate is $12$.

Such a query takes $2^{O(k)}\log^{2k}n$ time (\Cref{lem:range-tree}). Iterating over all $2^{O(1/\eps)}$ patterns and $k$ indices gives $\max_{v\in O}$ of the routed length in $\log^{O(1/\eps)}n$ time per vertex--separator incidence, and $O(n\log^{O(1/\eps)}n)$ time in total. We give the details in \Cref{sec:data-structure}.

Finally, computing the exact distances $\delta(u,s)$ and $\delta(t,v)$ by a shortest-path computation in $G$ for every separator set takes $\Omega(n)$ time per set, which is too expensive. We build Thorup's approximate distance oracle~\cite{thorup2004compact} once and use its estimates in place of exact distances. Since $P$ may have $\Theta(n)$ vertices, the portals of $u$ on $P$ are selected without scanning $P$. For vertices $s$ and $x$ in this order on $P$, the value $\delta(u,\cdot)-L(\cdot)$ drops from $s$ to $x$ by exactly $\delta(u,s)+\delta_P(s,x)-\delta(u,x)\ge0$. Hence this value is nonincreasing along $P$, and the first vertex at which it falls to at most a threshold is an in-portal for every later vertex whose value is less than $\eps\Delta/2$ below that threshold. Since the value varies by $O(\Delta)$, the portals are found by $O(1/\eps)$ binary searches along $P$, one for every multiple of $\eps\Delta/2$ in this range (\Cref{lem:global-portal-construction} and \Cref{fig:portal-search}). The search uses the oracle values, which need not be monotone, but each of them exceeds the monotone exact value by at most the oracle error, which suffices. Since the oracle never underestimates, the routed length computed from its estimates is at least $\delta(u,v)$. With estimates, the route through the successor may be longer than a route through a later out-portal, but only by two oracle errors, so the additive error increases only by $O(\eps\Delta)$ (\Cref{lem:ordering-approximate}).

\section{The Data Structure}\label{sec:data-structure}
In this section we present, for a single separator set $S$, a data structure that receives a vertex $u\in I_S$ and an interleaving pattern, and returns the largest estimated distance from $u$ to a vertex of $O_S$ that induces this pattern.

Throughout this section, fix a separator set $S\in\mathcal S$. A vertex lying on several paths of $S$ has a separate \emph{copy} on each of them. We index the copies by concatenating the paths of $S$, and let $\Ssin{u}$ (resp.\ $\Ssout{u}$) be the concatenation of the lists $\ssin{u}$ (resp.\ $\ssout{u}$) over $P\in S$. For two copies $x,y$, let $\delta_S(x,y)=\delta_P(x,y)$ if both lie on the same path $P\in S$, and $\delta_S(x,y)=\infty$ otherwise. Let $k=O(1/\eps)$ be the maximum length of these lists over the vertices of $I_S\cup O_S$. We give all lists length $k$ by repeating their last entry.

Let $\Pi_S$ be the set of interleaving patterns of two indexed lists $\iota_1,\ldots,\iota_k$ and $o_1,\ldots,o_k$ that preserve the order within each list and place $\iota_k$ before $o_k$. The $\iota_i$ stand for the entries of $\Ssin{u}$ and the $o_j$ for the entries of a destination's out-portal list, without recording vertex identities. Entries at the same position of $S$ are ordered with in-portal entries first, and entries of the same list by their list indices. Since the last vertex of every path is an out-portal, every in-portal has an out-portal at or after it on its path, so for every $u\in I_S$ and $v\in O_S$ the order of $\Ssin{u}$ and $\Ssout{v}$ along $S$ is a pattern in $\Pi_S$. For $u\in I_S$ and $\pi\in\Pi_S$, let $O_S(u,\pi)$ be the set of vertices $v\in O_S$ for which this order is $\pi$.

For $s\in\Ssin{u}$ and $v\in O_S$, let $N(s,v)$ be the first entry of $\Ssout{v}$ at or after $s$ on the path of $s$. For $\pi\in\Pi_S$, let $p_\pi(i)$ be the index of the first $o$-entry after $\iota_i$ in $\pi$. For every $v\in O_S(u,\pi)$, $N(\Ssin{u}[i],v)=\Ssout{v}[p_\pi(i)]$.

Every path of $S$ is a root path of the shortest-path tree $T$ of \Cref{lem:shortest-path-system}, rooted at $r$. For a vertex $x$ and for each of its copies, let $L(x)=\delta(r,x)$. Thus $\delta_S(x,y)=L(y)-L(x)$ whenever $\delta_S(x,y)<\infty$.
For every portal entry we store a distance estimate $\adist$ (\Cref{lem:global-portal-construction}) satisfying
\begin{equation}\label{eq:oracle-error}
    \delta(a,b)\le\adist(a,b)\le\delta(a,b)+\beta,
    \qquad \beta=\eps\hat\Delta/8.
\end{equation}
For $u\in I_S$, $v\in O_S$, $s\in\Ssin{u}$, and $t\in\Ssout{v}$, define the labels
\[
    a_u(s)=\adist(u,s)-L(s),\qquad b_v(t)=L(t)+\adist(t,v),
\]
and the \emph{routed estimate}
\[
    D_S(u,v)=\min_{i\in[k]}\Bigl(a_u(\Ssin{u}[i])+b_v\bigl(N(\Ssin{u}[i],v)\bigr)\Bigr).
\]

\begin{lemma}[Farthest-vertex data structure] \label{lem:data-structure-overall}
    In $O(|O_S|\log^{O(1/\eps)} n)$ time one can construct a data structure $\FarthestVertex(S)$ that,
    given $u\in I_S$ and $\pi\in \Pi_S$, returns $\max_{v\in O_S(u,\pi)}D_S(u,v)$ in $\log^{O(1/\eps)}n$ time.
\end{lemma}

\paragraph{Construction.}
For every $q\in [k]$, we embed every $v\in O_S$ as the point $\psi^q(v)\in\RR^{2k}$ with coordinates
\begin{itemize}
    \item $\psii{q}{j}(v)= b_v(\Ssout{v}[q]) - b_v(\Ssout{v}[j])$ for $1\le j\le k$, and
    \item $\psii{q}{k+j}(v)$ is the index of $\Ssout{v}[j]$ in $S$, for $1\le j\le k$,
\end{itemize}
with weight $b_v(\Ssout{v}[q])$.
$\FarthestVertex(S)$ consists of $k$ range trees $\OrthRange(1),\ldots,\OrthRange(k)$ of \Cref{lem:range-tree}, where $\OrthRange(q)$ is built on the weighted points $\psi^q(v)$, one for every $v\in O_S$. Construction takes $k\cdot 2^{O(k)}|O_S|\log^{2k-1}|O_S|=O(|O_S|\log^{O(1/\eps)}n)$ time.

\paragraph{Query.}
Given $u$ and $\pi$, for every $i\in[k]$ let $q=p_\pi(i)$, and query $\OrthRange(q)$ with the box defined by the following constraints, for all $j\in[k]$. Constraints~\eqref{eq:query-1} and~\eqref{eq:query-2} prescribe the successor of each in-portal according to $\pi$, and constraint~\eqref{eq:query-3} requires $i$ to attain the minimum in $D_S(u,v)$.
\begin{align}
    &\psii{q}{k+p_\pi(j)}(v) \ge \text{index of } \Ssin{u}[j]\text{ in $S$},    \label{eq:query-1} \\
    &\psii{q}{k+p_\pi(j)-1}(v) < \text{index of } \Ssin{u}[j]\text{ in $S$} \qquad \text{(only if $p_\pi(j)>1$)},   \label{eq:query-2}\\
    &\psii{q}{p_\pi(j)}(v) \leq a_u(\Ssin{u}[j]) - a_u(\Ssin{u}[i]). \label{eq:query-3}
\end{align}
Each constraint bounds one coordinate of $\psi^q(v)$ by a value that depends only on $u$, so the constraints define an axis-parallel box.
Let $M_i$ be the maximum weight in the box. The query returns $\max_{i\in[k]}\bigl(a_u(\Ssin{u}[i])+M_i\bigr)$.

\begin{proof}[Proof of \Cref{lem:data-structure-overall}]
    Fix $i$ and $q=p_\pi(i)$. We show that $v\in O_S$ lies in the box of $i$ if and only if $v\in O_S(u,\pi)$ and $i$ attains the minimum in $D_S(u,v)$.

    Constraints~\eqref{eq:query-1} and~\eqref{eq:query-2} state that $\Ssout{v}[p_\pi(j)-1]\prec\Ssin{u}[j]\preceq\Ssout{v}[p_\pi(j)]$ for every $j$. Since $p_\pi$ is nondecreasing and the out-portal list of $v$ is sorted, these relations place every in-portal entry in a specific gap of that list, which determines the interleaving. Hence they hold if and only if $v\in O_S(u,\pi)$.
    For $v\in O_S(u,\pi)$ we have $N(\Ssin{u}[j],v)=\Ssout{v}[p_\pi(j)]$ for every $j$. Hence, by the definition of $\psi^q$, constraint~\eqref{eq:query-3} for $j$ is equivalent to
    \[
        a_u(\Ssin{u}[i])+b_v(N(\Ssin{u}[i],v))\le a_u(\Ssin{u}[j])+b_v(N(\Ssin{u}[j],v)),
    \]
    and all of them together state that $i$ attains the minimum in $D_S(u,v)$.

    Consequently, for every $v$ in the box, $a_u(\Ssin{u}[i])+b_v(\Ssout{v}[q])=D_S(u,v)$, so $a_u(\Ssin{u}[i])+M_i$ is the maximum of $D_S(u,v)$ over the vertices of $O_S(u,\pi)$ for which $i$ is a minimizer. Taking the maximum over $i$ gives $\max_{v\in O_S(u,\pi)}D_S(u,v)$.

    A box query decomposes the box into at most $(2\log|O_S|+2)^{2k}$ canonical subsets and compares their stored maxima. Over all $k$ indices, a query takes $O(k^2+k(2\log|O_S|+2)^{2k})=\log^{O(1/\eps)}n$ time.
\end{proof}

\section{$(1+\eps)$-Approximation Algorithm for Eccentricities} \label{sec:algo}
In this section we prove our main result. Approximations to the diameter and the radius follow immediately from the eccentricities.

\Reminder{thm:eccentricities}

\paragraph{Preprocessing.}
We compute $\hat\Delta$ using~\Cref{lem:2-approx}, so $\frac{1}{2}\Delta\le\hat\Delta\le\Delta$.
We compute the separator system $\mathcal S$, the tree $T$, and the sets $I_S,O_S$ using~\Cref{lem:shortest-path-system}, together with $L(x)=\delta(r,x)$ for every $x\in V$.
Then, we build the data structure of \Cref{lem:global-portal-construction} once and use it to compute, for every $S\in\mathcal S$, the lists $\Ssin{u}$ for $u\in I_S$ and $\Ssout{v}$ for $v\in O_S$ with their estimates, which satisfy \Cref{lem:portal} and~\eqref{eq:oracle-error}. The index in $S$ of a portal is computed in $O(1)$ time from its depth in $T$.

\paragraph{The algorithm.}
After the preprocessing, the algorithm constructs $\FarthestVertex(S)$ for every $S\in \mathcal S$. Then, for every $S\in \mathcal S$, every $u\in I_S$, and every $\pi\in \Pi_S$, it sets $\hat{\ecc}(u)\gets\max(\hat{\ecc}(u), \FarthestVertex(S).\Query(u,\pi))$, where initially $\hat\ecc(u)=0$. See \Cref{alg:out-ecc}.

\begin{algorithm2e}[t]
\SetAlgoNoLine
\SetAlgoNoEnd
\caption{\codestyle{OutEccentricities$(G)$}} \label{alg:out-ecc}
\SetKwInOut{Input}{Input}
\Input{A strongly connected directed planar graph $G=(V,E,w)$ and $\eps\in(0,1]$}
For every $u\in V$: $\hat{\ecc}(u)\gets 0$ \\
\codestyle{Preprocess()} \\
\ForEach{$S\in \mathcal S$}{
    Construct $\FarthestVertex(S)$ \\
}
\ForEach{$S\in \mathcal S$, $u\in I_S$, $\pi\in \Pi_S$} {
    $\hat{\ecc}(u)\gets\max(\hat{\ecc}(u), \FarthestVertex(S).\Query(u, \pi))$
}
\Return{$\{\hat{\ecc}(u) \mid u\in V\}$}
\end{algorithm2e}

First, we bound the running time of the preprocessing.
\begin{lemma}[Preprocessing time] \label{lem:portal-time}
    The preprocessing takes $O(n\eps^{-2}\log^3 n\log(n/\eps))$ time.
\end{lemma}
\begin{proof}
    Computing $\hat\Delta$ and $L$ takes $O(n)$ time~\cite{henzinger1997faster}, and $\mathcal S$ takes $O(n\log n)$ time by \Cref{lem:shortest-path-system}. By \Cref{lem:shortest-path-system}, there are $O(n\log n)$ relevant vertex--separator incidences, each involving at most three paths. By \Cref{lem:global-portal-construction}, the shared oracle and the portal lists for all these incidences are computed in $O(n\eps^{-2}\log^3 n\log(n/\eps))$ time.
\end{proof}

The routed estimate never underestimates. For the pairs certified by the separator system, it overestimates by at most two portal errors $\eps\hat\Delta$ and two oracle errors $\beta$.
\begin{lemma}[Routed estimate] \label{lem:ordering-approximate}
    Let $S\in\mathcal S$, $u\in I_S$, and $v\in O_S$. Then $\delta(u,v)\le D_S(u,v)\le\delta(u,v)+3\eps\hat\Delta$.
\end{lemma}
\begin{proof}
    For the lower bound, let $s\in\Ssin{u}$ and $t=N(s,v)$. Then $t$ lies at or after $s$ on the same path of $S$, so $L(t)-L(s)=\delta_S(s,t)\ge\delta(s,t)$. By~\eqref{eq:oracle-error} and the triangle inequality,
    \[
        a_u(s)+b_v(t)=\adist(u,s)+\bigl(L(t)-L(s)\bigr)+\adist(t,v)\ge\delta(u,s)+\delta(s,t)+\delta(t,v)\ge\delta(u,v).
    \]

    For the upper bound, by \Cref{lem:shortest-path-system} there is a vertex $x\in P(u,v)\cap V(S)$. Fix a path $P\in S$ containing $x$ and consider the copy of $x$ on $P$. By \Cref{lem:portal}, there are $s\in\Ssin{u}$ and $s'\in\Ssout{v}$ on $P$ with
    \[
        \delta(u,s)+\delta_P(s,x)\le\delta(u,x)+\eps\hat\Delta,\qquad \delta_P(x,s')+\delta(s',v)\le\delta(x,v)+\eps\hat\Delta.
    \]
    Both left-hand sides are finite, so $s$ is at or before $x$ and $s'$ is at or after $x$ on $P$. Let $t=N(s,v)$. Since $s'$ is an out-portal of $v$ at or after $s$, the entry $t$ lies on $P$ at or after $s$ and not after $s'$. Hence $\delta(t,v)\le L(s')-L(t)+\delta(s',v)$. Using~\eqref{eq:oracle-error},
    \begin{align*}
        a_u(s)+b_v(t) &\le \delta(u,s)-L(s)+L(t)+\delta(t,v)+2\beta\\
        &\le \delta(u,s)+\bigl(L(s')-L(s)\bigr)+\delta(s',v)+2\beta\\
        &= \bigl(\delta(u,s)+\delta_P(s,x)\bigr)+\bigl(\delta_P(x,s')+\delta(s',v)\bigr)+2\beta\\
        &\le \delta(u,x)+\delta(x,v)+2\eps\hat\Delta+\eps\hat\Delta/4\le\delta(u,v)+3\eps\hat\Delta,
    \end{align*}
    where the last step uses that $x$ lies on a shortest path from $u$ to $v$.
\end{proof}

For every $u\in V$, let $\ecc^{\mathrm{out}}(u)=\max_{v\in V}\delta(u,v)$.
\begin{lemma}[Outgoing eccentricities] \label{lem:algo-correct}
    For every $u\in V$, $\ecc^{\mathrm{out}}(u) \leq \hat{\ecc}(u) \leq \ecc^{\mathrm{out}}(u) + 3\eps\hat\Delta$.
\end{lemma}
\begin{proof}
    By \Cref{lem:data-structure-overall}, and since the classes $O_S(u,\pi)$, $\pi\in\Pi_S$, cover $O_S$,
    \[
        \hat\ecc(u)=\max_{S\in\mathcal S(u)}\max_{v\in O_S}D_S(u,v).
    \]
    By \Cref{lem:ordering-approximate}, every $D_S(u,v)$ in this expression is at most $\delta(u,v)+3\eps\hat\Delta\le\ecc^{\mathrm{out}}(u)+3\eps\hat\Delta$, which proves the upper bound.
    For the lower bound, let $v$ satisfy $\delta(u,v)=\ecc^{\mathrm{out}}(u)$. By \Cref{lem:shortest-path-system} there is $S\in\mathcal S(u)$ with $v\in O_S$, and \Cref{lem:ordering-approximate} gives $\hat\ecc(u)\ge D_S(u,v)\ge\delta(u,v)$.
\end{proof}

\begin{lemma}[Running time] \label{lem:algo-time}
   \textnormal{$\codestyle{OutEccentricities}(G)$} takes $O(n\log^{O(1/\eps)}n)$ time.
\end{lemma}
\begin{proof}
    By \Cref{lem:portal-time}, the preprocessing takes $O(n\eps^{-2}\log^3 n\log(n/\eps))=O(n\log^{O(1/\eps)}n)$ time.
    By \Cref{lem:data-structure-overall,lem:shortest-path-system}, all data structures are constructed in $O(n\log^{O(1/\eps)}n)$ time, since $\sum_{S\in\mathcal S}|O_S|=O(n\log n)$.
    We have $|\Pi_S|\le\binom{2k}{k}\le 4^k=2^{O(1/\eps)}$. Hence, the queries take
    \[
        \sum_{S\in\mathcal S}|I_S|\cdot 2^{O(1/\eps)}\cdot\log^{O(1/\eps)}n=O(n\log n\cdot 2^{O(1/\eps)}\log^{O(1/\eps)}n)=O(n\log^{O(1/\eps)}n)
    \]
    time, using $\sum_{S\in\mathcal S}|I_S|=O(n\log n)$ from \Cref{lem:shortest-path-system}.
\end{proof}

\paragraph{The incoming direction.}
To approximate $\max_{v\in V}\delta(v,u)$, we run \Cref{alg:out-ecc} on the reverse graph $G^R$, in which $\delta_{G^R}(u,v)=\delta_G(v,u)$. Let $\hat\ecc^R(u)$ be its output, and return $E(u)=\max\{\hat\ecc(u),\hat\ecc^R(u)\}$. By \Cref{lem:algo-correct}, applied to both executions, and since $\hat\Delta\le\Delta\le2\ecc(u)$ (\Cref{lem:ecc-diameter}),
\[
    \ecc(u)\le E(u)\le\ecc(u)+3\eps\hat\Delta\le(1+6\eps)\ecc(u).
\]
\Cref{thm:eccentricities} follows by running both executions with parameter $\eps/6$, together with \Cref{lem:algo-time}.
Returning the maximum of the estimates gives a $(1+\eps)$-approximation of the diameter, which proves \Cref{thm:diameter}. Similarly, $\min_{u\in V}\ecc(u)\le\min_{u\in V}E(u)\le(1+\eps)\min_{u\in V}\ecc(u)$, so returning $\min_{u\in V}E(u)$ proves \Cref{cor:radius}.

\paragraph{Extensions.}
A vertex $c$ that attains $\min_{u\in V}E(u)$ satisfies $\ecc(c)\le(1+\eps)\min_{u\in V}\ecc(u)$, so it is an approximate center. Storing with every range maximum a vertex that attains it gives a pair $\pair{u,v}$ whose routed estimate is at least $\Delta$. By \Cref{lem:ordering-approximate} with parameter $\eps/6$, $\delta(u,v)\ge\Delta-\eps\hat\Delta/2\ge\Delta/(1+\eps)$.
The same algorithm approximates a two-sided version of the bichromatic diameter of~\cite{dalirrooyfard2019tight}, $\Delta_{A,B}=\max_{a\in A,b\in B}\max\{\delta(a,b),\delta(b,a)\}$ of a partition $V=A\mathbin{\dot\cup}B$. For every $S\in\mathcal S$ we build $\FarthestVertex(S)$ separately for $O_S\cap A$ and for $O_S\cap B$, and query every vertex in the structure of the other color. Two vertices of the same color are connected through a vertex of the other color, so $\Delta\le2\Delta_{A,B}$ and the additive error $3\eps\hat\Delta$ is at most $6\eps\Delta_{A,B}$.

\bibliography{bibliography}

\section*{AI Statement}
ChatGPT and Claude were used to assist with drafting and revising the paper, checking mathematical arguments, and verifying references.
The authors take full responsibility for the final content, including all results, proofs, and citations.

\clearpage
\appendix

\section{A Shortest-Path Separator System with the In-Out Property}
\label{sec:in-out-separators}

In this section we prove the in-out shortest-path separator system of \Cref{lem:shortest-path-system}. We first construct the hierarchy and show that it covers every ordered pair of distinct vertices (\Cref{lem:active-set-hierarchy}). We then use a reduced cotree at each hierarchy node to bound the total construction time (\Cref{lem:reduction,lem:active-hierarchy-construction}).

\Reminder{lem:shortest-path-system}

Let $\bar G$ be a triangulated plane augmentation of the undirected planar graph obtained from $G$ by ignoring the
directions of its edges and merging antiparallel arcs into one edge.  A partition that separates $\bar G$ also separates $G$.
Fix a vertex $r\in V$ and a directed shortest-path
tree $T$ of $G$ rooted at $r$, with every tree path directed away from $r$.
After ignoring its edge
directions, $T$ is a spanning tree of $\bar G$.  Whenever $P(r,z)$ appears below, it denotes
the $r$-to-$z$ path in $T$.
We use the following standard vertex-cost form of the
rooted-path separator lemma.  
\begin{lemma}[Vertex-cost rooted-path separator
\cite{lipton1979separator,thorup2004compact}]
\label{lem:vertex-cost-rooted-path-separator}
Let $c:V\rightarrow\RR_{\geq 0}$ be a vertex-cost function, and write
$c(X)=\sum_{x\in X}c(x)$.  There are three
vertices $z_1,z_2,z_3$ such that, if
\[
    \Sigma=\{P(r,z_i):i\in[3]\},
    \qquad
    V_\Sigma=V(\Sigma)=\bigcup_{i=1}^{3}V(P(r,z_i)),
\]
then there is a partition
\[
    V\setminus V_\Sigma
    =
    R_1\mathbin{\dot\cup}R_2\mathbin{\dot\cup}R_3
\]
such that
\begin{enumerate}
    \item no edge of $\bar G$ has endpoints in two distinct parts;
    \item $c(R_i)\leq c(V)/2$ for every $i\in[3]$.
\end{enumerate}
The paths and the sides can be found using the algorithmic rooted-path
separator construction of Thorup~\cite[Section~2, Lemma~2.3]{thorup2004compact}.
\end{lemma}

We call the parts
$R_1,R_2,R_3$ the \emph{separator sides}.

\paragraph{The active-set hierarchy.}
We define a tree hierarchy of vertex sets.
At every node of this hierarchy, we apply \Cref{lem:vertex-cost-rooted-path-separator} with a different cost function.

We recursively construct a rooted hierarchy $\mathcal H$.  Every hierarchy
node $\tau$ has a nonempty \emph{active set} $A_\tau\subseteq V$.  The root $\rho$
has $A_\rho=V$.

Let $\tau$ be a hierarchy node, starting with $\rho$.  Define
\[
    c_\tau(x)=
    \begin{cases}
        1 & x\in A_\tau,\\
        0 & x\notin A_\tau.
    \end{cases}
\]
We apply \Cref{lem:vertex-cost-rooted-path-separator} to $\bar G$, using the
fixed shortest-path tree $T$ and the cost function $c_\tau$.  The
implementation and running-time analysis are given in the proof of
\Cref{lem:active-hierarchy-construction}.
Let $P_{\tau,i}=P(r,z_{\tau,i})$, for $i\in[3]$, be the resulting paths,
and define
\[
    S_\tau=\{P_{\tau,i}:i\in[3]\},
    \qquad
    V(S_\tau)=\bigcup_{i=1}^{3}V(P_{\tau,i}).
\]
Let $R_{\tau,1},R_{\tau,2},R_{\tau,3}$ be the separator sides
returned by \Cref{lem:vertex-cost-rooted-path-separator}.  Define
\[
    B_\tau=A_\tau\cap V(S_\tau),
    \qquad
    A_{\tau,i}=A_\tau\cap R_{\tau,i}
    \quad (i\in[3]).
\]
For every $i$ with $A_{\tau,i}\neq\emptyset$, create a child $\tau_i$ of
$\tau$ and set $A_{\tau_i}=A_{\tau,i}$.  Thus the vertices in $B_\tau$ are
removed from the active sets below $\tau$.

\begin{figure}[t]
    \centering
    \begin{tikzpicture}[
        scale=0.88,
        vertex/.style={circle,draw=black,fill=white,inner sep=1.6pt},
        active/.style={circle,draw=blue!70!black,fill=blue!20,inner sep=2pt},
        candidate/.style={circle,draw=green!50!black,fill=green!25,inner sep=2pt},
        boundary/.style={circle,draw=red!70!black,fill=red!25,inner sep=2pt},
        >=stealth
    ]
        \draw[rounded corners=15pt,draw=blue!55,dashed,fill=blue!5]
            (-2.25,-0.05) rectangle (-0.35,1.45);
        \draw[rounded corners=15pt,draw=green!50!black,dashed,fill=green!5]
            (0.35,-0.05) rectangle (2.15,1.45);
        \draw[rounded corners=15pt,draw=orange!70!black,dashed,fill=orange!6]
            (2.85,-0.05) rectangle (4.45,1.45);

        \node[blue!70!black] at (-1.30,1.65) {$R_{\tau,1}$};
        \node[green!45!black] at (1.25,1.65) {$R_{\tau,2}$};
        \node[orange!70!black] at (3.65,1.65) {$R_{\tau,3}$};

        \coordinate (r) at (0,-1.65);
        \draw[very thick,red!75!black,->]
            (r) .. controls (-2.35,-1.15) and (-2.75,0.15) .. (-2.75,2.25)
            node[above left,black] {$z_1$};
        \draw[very thick,red!75!black,->]
            (r) .. controls (0,-0.45) and (0,0.75) .. (0,2.55)
            node[above,black] {$z_2$};
        \draw[very thick,red!75!black,->]
            (r) .. controls (2.25,-1.10) and (2.55,0.20) .. (2.65,2.25)
            node[above right,black] {$z_3$};
        \draw[gray!80,thick]
            (-2.75,2.25) -- (0,2.55) -- (2.65,2.25)
            .. controls (2.65,4.65) and (-2.75,4.65) .. (-2.75,2.25);
        \node at (0,3.20) {$g$};
        \node[vertex,label=below:$r$] at (r) {};
        \node[red!75!black] at (-0.20,-1.05) {$S_\tau$};

        \node[active,label=left:$u$] (u) at (-1.80,0.55) {};
        \node[active,label=below:$x$] (x) at (-0.85,0.35) {};
        \node[candidate,label=right:$v_2$] (v2) at (0.85,0.75) {};
        \node[candidate,label=right:$v_3$] (v3) at (3.45,0.65) {};
        \node[boundary,label=below right:$b$] (b) at (0,0.95) {};

        \draw[densely dashed,green!45!black,thick]
            (u) .. controls (-1.20,1.20) and (-0.45,1.15) .. (b);
        \draw[->,densely dashed,green!45!black,thick]
            (b) .. controls (0.30,0.80) and (0.50,0.75) .. (v2);
        \draw[->,densely dashed,green!45!black,thick]
            (u) .. controls (0.15,-0.35) and (2.20,-0.25) .. (v3);
    \end{tikzpicture}
    \caption{One step of the active-set hierarchy. The separator $S_\tau$ (red) consists of the tree paths from $r$ to the three vertices of the face $g$. The set $S_{\tau,1}$ has in-set $A_{\tau,1}\cup B_\tau$ and out-set $A_\tau\setminus A_{\tau,1}$. For this set, $u\in A_{\tau,1}$ is a source for $b$, $v_2$, and $v_3$, and the pair $\pair{u,x}$ is passed to the child $\tau_1$.}
    \label{fig:in-out-step}
\end{figure}
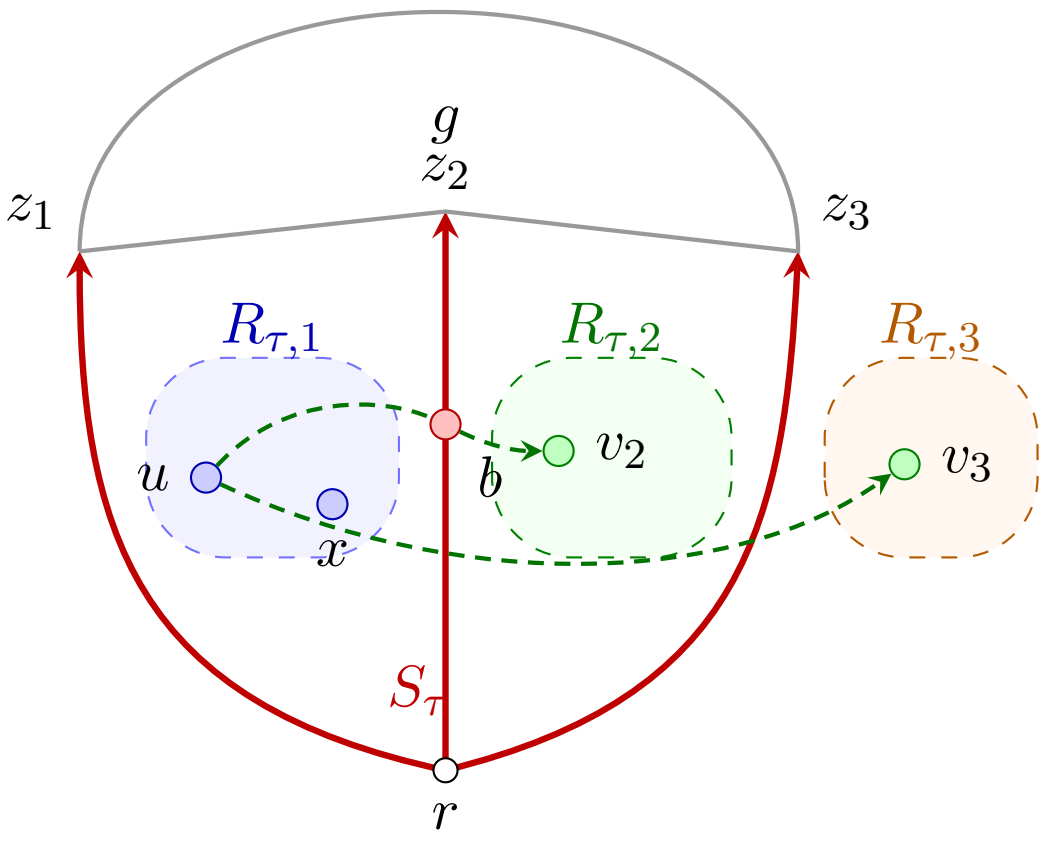

For every $i\in[3]$, add a copy $S_{\tau,i}$ of
$S_\tau$ to $\mathcal S$ and define
\[
    I_{S_{\tau,i}}=A_{\tau,i} \cup B_\tau,
    \qquad
    O_{S_{\tau,i}}=A_\tau\setminus A_{\tau,i}.
\]
A vertex of side $i$ is a source for the destinations outside side $i$, and the pairs inside side $i$ are passed to the child. Vertices of $B_\tau$ are not active below $\tau$, so they are sources and destinations in all three copies.
See \Cref{fig:in-out-step} for an illustration of the separator and our definitions.

Next, we prove some properties of our hierarchy.
\begin{lemma}[Properties of the hierarchy]
\label{lem:active-set-hierarchy}
The hierarchy described above has the following properties:
\begin{enumerate}
    \item Every child active set satisfies
    $|A_{\tau,j}|\leq |A_\tau|/2$.
    Consequently, the hierarchy has depth $O(\log n)$,
    $\sum_{\tau\in V(\mathcal H)}|A_\tau|=O(n\log n)$, and it contains
    $O(n)$ hierarchy nodes. \label{prop:1}
    \item For every distinct $u,v\in V$, there is a node $\tau$ and an indexed set $S_{\tau,i}$ such that $u\in I_{S_{\tau,i}}$ and $v\in O_{S_{\tau,i}}$.\label{prop:2}
    \item For every indexed set $S_{\tau,i}$, every
    $u\in I_{S_{\tau,i}}$, and every $v\in O_{S_{\tau,i}}$, it holds that
    $P(u,v)\cap V(S_\tau)\neq\emptyset$ and
    $P(v,u)\cap V(S_\tau)\neq\emptyset$.\label{prop:3}
\end{enumerate}
\end{lemma}
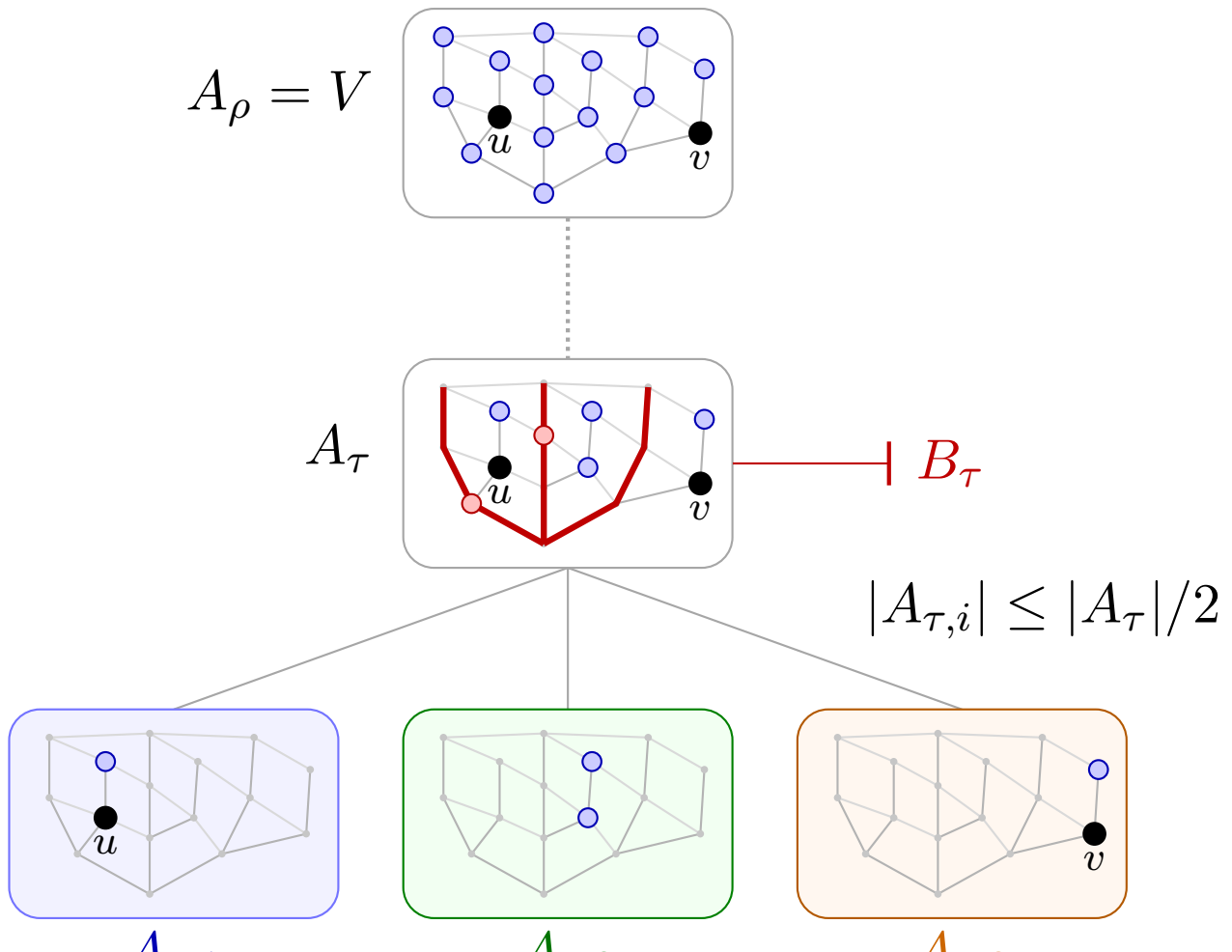
\begin{figure}[t]
    \centering
    \newcommand{\hgraph}{%
        \coordinate (r) at (1.5,0);
        \coordinate (a1) at (0.6,0.5);  \coordinate (a2) at (0.25,1.2);  \coordinate (a3) at (0.25,1.95);
        \coordinate (b1) at (1.5,0.7);  \coordinate (b2) at (1.5,1.35);  \coordinate (b3) at (1.5,2.0);
        \coordinate (c1) at (2.4,0.5);  \coordinate (c2) at (2.75,1.2);  \coordinate (c3) at (2.8,1.95);
        \coordinate (q1) at (0.95,0.95); \coordinate (q2) at (0.95,1.65);
        \coordinate (q3) at (2.05,0.95); \coordinate (q4) at (2.1,1.65);
        \coordinate (q5) at (3.45,0.75); \coordinate (q6) at (3.5,1.55);
        \draw[gray!30] (a2)--(q1) (q2)--(a3) (q2)--(b2) (q1)--(b1) (q3)--(b2)
            (q4)--(b3) (q4)--(c2) (q3)--(c1) (q5)--(c2) (q6)--(c3)
            (a3)--(b3)--(c3);
        \draw[gray!60] (r)--(a1)--(a2)--(a3) (r)--(b1)--(b2)--(b3)
            (r)--(c1)--(c2)--(c3) (a1)--(q1)--(q2) (b1)--(q3)--(q4)
            (c1)--(q5)--(q6);
        \foreach \p in {r,a1,a2,a3,b1,b2,b3,c1,c2,c3,q1,q2,q3,q4,q5,q6}
            \fill[gray!45] (\p) circle (1.3pt);
    }
    \begin{tikzpicture}[
        act/.style={circle,draw=blue!70!black,fill=blue!20,inner sep=1.3pt},
        bnd/.style={circle,draw=red!70!black,fill=red!25,inner sep=1.3pt},
        pr/.style={circle,draw=black,fill=black,inner sep=1.5pt},
        box/.style={rounded corners=6pt,draw=gray!70}
    ]
        \begin{scope}[shift={(-0.99,4.8)},scale=0.55]
            \draw[box] (-0.25,-0.3) rectangle (3.85,2.3);
            \hgraph
            \foreach \p in {r,a1,a2,a3,b1,b2,b3,c1,c2,c3,q2,q3,q4,q6}
                \node[act] at (\p) {};
            \node[pr,label={[inner sep=1pt]below:\scriptsize$u$}] at (q1) {};
            \node[pr,label={[inner sep=1pt]below:\scriptsize$v$}] at (q5) {};
        \end{scope}
        \node[left] at (-1.2,5.45) {$A_\rho=V$};

        \begin{scope}[shift={(-0.99,2.4)},scale=0.55]
            \draw[box] (-0.25,-0.3) rectangle (3.85,2.3);
            \hgraph
            \draw[very thick,red!75!black] (r)--(a1)--(a2)--(a3)
                (r)--(b1)--(b2)--(b3) (r)--(c1)--(c2)--(c3);
            \foreach \p in {q2,q3,q4,q6} \node[act] at (\p) {};
            \foreach \p in {a1,b2} \node[bnd] at (\p) {};
            \node[pr,label={[inner sep=1pt]below:\scriptsize$u$}] at (q1) {};
            \node[pr,label={[inner sep=1pt]below:\scriptsize$v$}] at (q5) {};
        \end{scope}
        \node[left] at (-1.2,3.05) {$A_\tau$};

        \foreach \cx/\col/\i in {-2.7/blue!70!black/1, 0/green!45!black/2, 2.7/orange!80!black/3} {
            \draw[gray!70] (0,2.235) -- (\cx,1.265);
            \node[\col] at (\cx,-0.45) {$A_{\tau,\i}$};
        }
        \begin{scope}[shift={(-3.69,0)},scale=0.55]
            \draw[box,draw=blue!55,fill=blue!5] (-0.25,-0.3) rectangle (3.85,2.3);
            \hgraph
            \node[act] at (q2) {};
            \node[pr,label={[inner sep=1pt]below:\scriptsize$u$}] at (q1) {};
        \end{scope}
        \begin{scope}[shift={(-0.99,0)},scale=0.55]
            \draw[box,draw=green!50!black,fill=green!5] (-0.25,-0.3) rectangle (3.85,2.3);
            \hgraph
            \foreach \p in {q3,q4} \node[act] at (\p) {};
        \end{scope}
        \begin{scope}[shift={(1.71,0)},scale=0.55]
            \draw[box,draw=orange!70!black,fill=orange!6] (-0.25,-0.3) rectangle (3.85,2.3);
            \hgraph
            \node[act] at (q6) {};
            \node[pr,label={[inner sep=1pt]below:\scriptsize$v$}] at (q5) {};
        \end{scope}

        \draw[gray!70,densely dotted,thick] (0,4.635) -- (0,3.665);
        \draw[red!75!black] (1.13,2.95) -- (2.2,2.95);
        \draw[red!75!black,thick] (2.2,2.8) -- (2.2,3.1);
        \node[right,red!75!black] at (2.25,2.95) {$B_\tau$};
        \node[right] at (1.9,1.95) {$|A_{\tau,i}|\le|A_\tau|/2$};

    \end{tikzpicture}
    \caption{The hierarchy on the fixed graph and tree. The pair $u,v$ is first separated at $\tau$.}
    \label{fig:in-out-hierarchy}
\end{figure}

\begin{proof}
We start with Property \ref{prop:1}.  Since $c_\tau(V)=|A_\tau|$,
\Cref{lem:vertex-cost-rooted-path-separator} gives
$|A_{\tau,i}|=c_\tau(R_{\tau,i})\leq|A_\tau|/2$ for every $i\in[3]$, so every child of $\tau$ contains at most $|A_\tau|/2$ vertices.  Consequently,
the size of the active set halves along every root-to-node path of
$\mathcal H$, and the depth of $\mathcal H$ is at most $\log_2 n$. Active sets of the same level are disjoint, so each vertex appears in at most $\log_2 n+1$ active sets, and $\sum_{\tau\in V(\mathcal H)}|A_\tau|=O(n\log n)$.

It remains to bound the number of hierarchy nodes. The leaves of $\mathcal H$ have disjoint nonempty active sets, so $\mathcal H$ has at most $n$ leaves, and therefore fewer than $n$ nodes with at least two children. Consider a node $\tau$ with exactly one child $\sigma$. Then $A_\tau=A_\sigma\mathbin{\dot\cup}B_\tau$, and since $|A_\sigma|\leq|A_\tau|/2$ we get $|B_\tau|\geq1$. The sets $B_\tau$ are pairwise disjoint, since active sets of incomparable nodes are disjoint and a vertex of $B_\tau$ is in no active set below $\tau$. Hence $\mathcal H$ has at most $n$ nodes with exactly one child, and $O(n)$ nodes in total.

Next we prove Property \ref{prop:2}. Let $u,v\in V$ be two distinct vertices, and let $\tau$ be the deepest hierarchy node such that $u,v\in A_\tau$ (\Cref{fig:in-out-hierarchy}).

Recall that $A_\tau$ is the disjoint union of $B_\tau$, $A_{\tau,1}$, $A_{\tau,2}$, and $A_{\tau,3}$, and that no single $A_{\tau,i}$ contains both $u$ and $v$, since otherwise the corresponding child of $\tau$ would contradict the choice of $\tau$.  We choose the index $i$ as follows.  If $u\in A_{\tau,i}$ for some $i\in[3]$, we take that index. Otherwise $u\in B_\tau$, and we take any $i\in [3]$ with $v\notin A_{\tau,i}$.  In both cases $u\in A_{\tau,i}\cup B_\tau = I_{S_{\tau,i}}$ and $v\in A_\tau\setminus A_{\tau,i} = O_{S_{\tau,i}}$, as required.

Finally, we prove Property \ref{prop:3}.  Let $S=S_{\tau,i}$, let
$u\in I_S$, and let $v\in O_S$.  If either endpoint lies in
$B_\tau\subseteq V(S_\tau)$, both paths meet $V(S_\tau)$ at that endpoint.
Otherwise, $u\in A_{\tau,i}$, whereas
$v\in O_S$ lies in a different side block $A_{\tau,i'}$.
Since $A_{\tau,i}\subseteq R_{\tau,i}$ and $A_{\tau,i'}\subseteq R_{\tau,i'}$, the vertices $u$ and $v$ lie in distinct separator sides.
From \Cref{lem:vertex-cost-rooted-path-separator} we have that no edge of $\bar G$ joins distinct separator sides, and therefore every path between
$u$ and $v$ in either direction meets $V(S_\tau)$.  In particular, this
holds for $P(u,v)$ and $P(v,u)$.
\end{proof}

\paragraph{Efficient construction.}
Next, we bound the construction time of the separator system.  A direct
application of \Cref{lem:vertex-cost-rooted-path-separator} at a hierarchy node
takes $\Omega(n)$ time, which is too slow since $\mathcal H$ may have
$\Omega(n)$ nodes. Instead, we run the construction on a tree of size $O(|A_\tau|)$, obtained by shrinking the cotree defined next.

Let $\bar G^*$ be the planar dual of $\bar G$, and let $K$ be the \emph{cotree}
of $T$, that is, the subgraph of $\bar G^*$ formed by the duals of the edges of
$\bar G$ that are not in $T$.  Since $T$ is a spanning tree of the connected
plane graph $\bar G$, the cotree $K$ is a spanning tree of
$\bar G^*$~\cite[Section~2]{thorup2004compact}: its nodes are the faces of
$\bar G$, and, as $\bar G$ is triangulated and hence every face is incident to
exactly three edges, $K$ has maximum degree three.  Fix for every vertex
$x\in V$ an incident face $f(x)$, and transfer the costs from the vertices to
the nodes of $K$ by setting $c(g)=c(\{x\in V : f(x)=g\})$ for a node $g$ of
$K$, and $c(X)=\sum_{g\in X}c(g)$ for a set $X$ of nodes.

The construction of Thorup~\cite[Section~2, Lemma~2.3]{thorup2004compact} that
underlies \Cref{lem:vertex-cost-rooted-path-separator} reads the cost function
only through these node costs.  It computes $K$ and $f$ from $\bar G$ and $T$ in
$O(n)$ time, independently of $c$, and then selects a node $g$ of $K$ such that
every component of $K-g$ has cost at most $c(V(K))/2$.  The vertices
$z_1,z_2,z_3$ are the three vertices on the boundary of the face $g$, and the
side $R_i$ consists of the vertices $x\notin V_\Sigma$ whose face $f(x)$ lies in the
$i$-th component of $K-g$. In particular there are at most three sides, since
$g$ has at most three neighbors in $K$.  Hence the only step whose running time
is not $O(1)$ per vertex is the selection of $g$, which is done by a single
traversal of $K$. This traversal takes $\Omega(n)$ time even when only $|A_\tau|$ vertices have positive cost.

The sides are separated for the following reason. Every edge of $K$ incident to $g$ is dual to a non-tree edge $ab$ on the boundary of $g$. The vertices of the cycle formed by $ab$ and the path of $T$ between $a$ and $b$ lie on $P(r,a)\cup P(r,b)$, so they are all in $V_\Sigma$, and the component of $K-g$ attached to $g$ through this edge consists of the faces on the side of the cycle that does not contain $g$. A vertex $x\notin V_\Sigma$ is on none of these cycles, so all the faces incident to $x$ lie in the same component of $K-g$, and the side of $x$ does not depend on the choice of $f(x)$. Similarly, an edge of $\bar G$ whose endpoints are both outside $V_\Sigma$ crosses none of these cycles, so its endpoints lie in the same side.

At a node $\tau$, we obtain this tree from $K$ by discarding the parts that carry no cost, i.e., removing all nodes of degree at most $2$ that have cost zero.

A $c$-reduction tree $K'$ is a topological minor of $K$, obtained by repeatedly
deleting or suppressing nodes of degree at most two whose cost is zero, in which no
node of degree at most two has cost zero. Suppressing a node of degree two replaces it and its two edges by one edge between its two neighbors. Since every deleted node has
cost zero, we have that $c(V(K'))=c(V(K))$, and it contains every node of positive cost.

$\codestyle{Reduce}(K,c)$ is a straightforward procedure that takes a tree $K$ and a cost function $c$ and returns a $c$-reduction tree $K'$. $\codestyle{Reduce}(K,c)$ takes $O(|V(K)|)$ time by maintaining a queue of the nodes of degree at most $2$ and cost $0$, and repeatedly deleting (or suppressing) them, updating the degrees of their neighbors. 
Next, we prove the following useful properties of $\codestyle{Reduce}(K,c)$.

\begin{lemma}
\label{lem:reduction}
Let $K'=\codestyle{Reduce}(K,c)$, and let $g$ be a node of $K'$.  Then
\begin{enumerate}
    \item If every component of $K'-g$ has cost at most $c(V(K'))/2$, then every component of $K-g$ has cost at most
    $c(V(K))/2$;\label{prop:red-centroid}
    \item $|V(K')|\leq 2|\{h\in V(K)\mid c(h)>0\}|$.\label{prop:red-size}
\end{enumerate}
\end{lemma}

\begin{proof}
For Property~\ref{prop:red-centroid}, every node of $K$ that is not in $K'$ has cost zero. Hence every component of $K-g$ is obtained from a component of $K'-g$ by adding zero-cost nodes, or has cost zero. Since $c(V(K))=c(V(K'))$, the property follows.

For Property~\ref{prop:red-size}, note that $\codestyle{Reduce}$ terminates only when
every node of degree at most $2$ has positive cost.  In particular, every leaf of
$K'$ has positive cost.  Hence every zero-cost node of $K'$ has degree at least
$3$, and in any tree the number of nodes of degree at least $3$ is smaller than
the number of leaves.  Therefore
\[
    |\{h\in V(K')\mid c(h)=0\}| \;\le\; |\{h\in V(K') \mid h \text{ is a leaf}\}|
    \;\le\; |\{h\in V(K)\mid c(h)>0\}|,
\]
and since $V(K')$ is the disjoint union of its zero-cost and positive-cost nodes,
$|V(K')|\le 2|\{h\in V(K)\mid c(h)>0\}|$.
\end{proof}

Next, we bound the construction time.
\begin{lemma}[Construction time]
\label{lem:active-hierarchy-construction}
The hierarchy $\mathcal H$, the collection $\mathcal S$ together with its
separator paths, and all the in-sets and out-sets, can be computed in
$O(n\log n)$ time.
\end{lemma}

\begin{proof}
Computing $\bar G$, $T$~\cite{henzinger1997faster}, and the cotree $K$ takes $O(n)$ time. In addition, we compute a preorder numbering of $T$ together with the subtree interval of every vertex, in $O(n)$ time, so that one can test in $O(1)$ time whether a vertex $x$ is an ancestor of a vertex $z$ in $T$, i.e., whether $x\in V(P(r,z))$. Each separator path $P(r,z)$ is represented implicitly by its last vertex $z$.

Let $\tau$ be a hierarchy node. Next, we show that processing $\tau$  and creating its children takes $O(|A_\tau|)$ time.
The cost that $c_\tau$ transfers to a node $g$
of $K$ is $c_\tau(g)=|\{x\in A_\tau : f(x)=g\}|$, so the total cost is
$|A_\tau|$ and at most $|A_\tau|$ nodes have positive cost.

The recursion keeps
at $\tau$ a $c_\tau$-reduction $K_\tau$ of $K$.  Note that $K_\tau$ is obtained by
reducing the tree $K_\sigma$ of the parent $\sigma$ of $\tau$ rather than $K$ itself.
This is indeed a $c_\tau$-reduction of $K$, since $A_\tau\subseteq A_\sigma$ and
therefore every node deleted or suppressed along the way, having zero
$c_\sigma$-cost, also has zero $c_\tau$-cost.  Hence $K_\tau$ is obtained from $K$ by
repeatedly deleting or suppressing nodes of zero $c_\tau$-cost, and by
Property~\ref{prop:red-size} of \Cref{lem:reduction} it has
$|V(K_\tau)|\leq2|A_\tau|$ nodes.

By Property~\ref{prop:red-centroid} of \Cref{lem:reduction}, a node $g$ that is balanced in $K_\tau$ is also balanced in $K$. The two partitions are not identical, but every vertex of $A_\tau\setminus B_\tau$ lies in the component of $K-g$ that lifts the component of $K_\tau-g$ containing its face. Hence we may run
the selection step on $K_\tau$ rather than on $K$, in $O(|A_\tau|)$ time.

Constructing $B_\tau=A_\tau\cap V(S_\tau)$ takes $O(|A_\tau|)$ time, by testing for every $x\in A_\tau$ whether $x$ is an ancestor of one of $z_{\tau,1},z_{\tau,2},z_{\tau,3}$ in $T$. Similarly, the side containing a vertex $x\in A_\tau\setminus B_\tau$ is determined in $O(1)$ time from the component of $K_\tau-g$ containing $f(x)$. 
The six sets $I_{S_{\tau,i}}$ and $O_{S_{\tau,i}}$ are
then listed from their definitions in $O(|A_\tau|)$ time, as each of them is a
subset of $A_\tau$.

It remains to pass a tree to every child $\tau_j$ of $\tau$.  
For each of the at most three children, we set the costs of a copy of $K_\tau$ to $c_{\tau_j}$ by one scan of $A_{\tau_j}$, and computing $K_{\tau_j}=\codestyle{Reduce}(K_\tau,c_{\tau_j})$ on this copy takes $O(|V(K_\tau)|)=O(|A_\tau|)$ time.

Overall, we get that processing a hierarchy node $\tau$ takes $O(|A_\tau|)$ time, and by
\Cref{lem:active-set-hierarchy} the total time over the hierarchy is
$\sum_{\tau\in V(\mathcal H)}O(|A_\tau|)=O(n\log n)$, as required.
\end{proof}

Next, we complete the section by proving \Cref{lem:shortest-path-system}.
\begin{proof}[Proof of \Cref{lem:shortest-path-system}]
The construction above defines the indexed collection $\mathcal S$ and the
sets $I_S$ and $O_S$ for every $S\in\mathcal S$.  Each $S$ is a set
of at most three directed shortest paths by
\Cref{lem:vertex-cost-rooted-path-separator}.

Fix two distinct vertices $u,v$.  By
\Cref{lem:active-set-hierarchy}, there is an $S\in\mathcal S(u)$ such that
$v\in O_S$.  This proves the first property of
\Cref{lem:shortest-path-system}.  The second property is exactly the last property of
\Cref{lem:active-set-hierarchy}.

For a fixed vertex $u$, only the three copies created at a hierarchy node
$\tau$ whose active set contains $u$ may have $u$ in their in-set, so every
such node contributes $O(1)$ members to $\mathcal S(u)$.  The nodes whose
active set contains $u$ lie on a single root-to-node path of $\mathcal H$,
since active sets of incomparable nodes are disjoint, and the hierarchy has
depth $O(\log n)$.  Hence $|\mathcal S(u)|=O(\log n)$.  Every in-set and out-set
created at $\tau$ is a subset of $A_\tau$, and three
separator sets are created at $\tau$, so by
\Cref{lem:active-set-hierarchy},
\[
    \sum_{S\in\mathcal S}|I_S|
    +
    \sum_{S\in\mathcal S}|O_S|
    \leq
    6\sum_{\tau\in V(\mathcal H)}|A_\tau|
    =
    O(n\log n),
\]
which proves the third
property of \Cref{lem:shortest-path-system}.

Every hierarchy node creates three sets, so $|\mathcal S|=O(n)$ by \Cref{lem:active-set-hierarchy}, and all the paths are root-to-vertex paths of $T$ by construction.
The claimed construction time, and the representations of $\mathcal S$, of the
in-sets, and of the out-sets, follow from
\Cref{lem:active-hierarchy-construction}.
\end{proof}

\section{Shared Preprocessing for Portal Construction}
\label{sec:portal-preprocessing}

In this section we explain how to compute the portals of \Cref{lem:portal} efficiently for all the paths of the in-out separator system. Thorup~\cite{thorup2004compact} computes portals inside his recursive decomposition, where every separator path is processed in a subgraph that shrinks with the recursion. Our separator system keeps $G$ fixed, so computing distances in $G$ for every separator set takes $\Omega(n)$ time per hierarchy node, although there are only $O(n\log n)$ relevant vertex--separator incidences. We instead build one approximate distance oracle for $G$ and use it to select the portals on implicitly represented tree paths.

\begin{lemma}[Portal construction]\label{lem:global-portal-construction}
Let $\hat\Delta$ satisfy $\hat\Delta\le\Delta\le2\hat\Delta$, and let $T$ be a shortest-path tree of $G$ rooted at $r$. In $O(n\eps^{-2}\log^3 n\log(n/\eps))$ time one can construct a data structure with the following properties.
\begin{enumerate}
    \item A query returns a value $\adist(a,b)$ such that
    \[
        \delta(a,b)\le\adist(a,b)\le\delta(a,b)+\beta,
        \qquad \beta=\eps\hat\Delta/8,
    \]
    in $Q=O(\eps^{-1}+\log\log(n/\eps))$ time.
    \item For a specified vertex $u$ and a specified tree path $P=P(r,z)$, portal lists satisfying \Cref{lem:portal}, together with their distance estimates, can be computed in $O(\eps^{-1}\log n\,(\log n+Q))$ time. The lists include both endpoints of $P$.
\end{enumerate}
In particular, all the portal lists for the $O(n\log n)$ relevant incidences of our separator system can be computed within the stated preprocessing bound.
\end{lemma}

\begin{proof}
We first build an approximate distance oracle, then select the in-portals, and finally select the out-portals and bound the total time.

\paragraph{The oracle.}
To apply Thorup's oracle, we convert the weights to nonnegative integers. Remove arcs of weight greater than $2\hat\Delta$. This preserves every distance: a simple shortest path has length at most $\Delta\le2\hat\Delta$, so none of its nonnegative-weight arcs is removed. Set
\[
    \lambda=\frac{\eps\hat\Delta}{32n},
    \qquad w'(e)=\ceil{w(e)/\lambda}.
\]
The resulting planar digraph has integer weights in $\{0,\ldots,N\}$, with $N\le\ceil{64n/\eps}$. Writing $\delta'$ for its distances, rounding each arc on a simple shortest path gives
\[
    \delta(a,b)\le\lambda\delta'(a,b)
    \le\delta(a,b)+n\lambda.
\]
Build Thorup's deterministic directed distance oracle~\cite[Theorem~3.16]{thorup2004compact} with stretch $1+\rho$, where $\rho=\eps/64$. Its preprocessing time is $O(n\rho^{-2}\log^3 n\log(nN))$, and its query time is $O(\rho^{-1}+\log\log(nN))$. Multiply its answers by $\lambda$ to define $\adist$. Then
\begin{align*}
    \delta(a,b)\le\adist(a,b)
    &\le(1+\rho)(\delta(a,b)+n\lambda)\\
    &\le\delta(a,b)+\frac{\eps\hat\Delta}{32}
       +\frac{\eps}{64}\left(2\hat\Delta+
                                  \frac{\eps\hat\Delta}{32}\right)
     \le\delta(a,b)+\beta.
\end{align*}

\paragraph{In-portals.}
Precompute depths, root distances, and binary-lifting pointers for $T$ in $O(n\log n)$ time. They give access to any indexed vertex of $P=P(r,z)$ in $O(\log n)$ time. Write $m=|P|$ and $L_i=\delta(r,P[i])$. For a fixed $u$, define
\[
    f_i=\delta(u,P[i])-L_i,
    \qquad \widetilde f_i=\adist(u,P[i])-L_i.
\]
The triangle inequality along $P$ shows that $f_i$ is nonincreasing. Also,
\[
    -2\hat\Delta\le f_i\le2\hat\Delta,
    \qquad f_i\le\widetilde f_i\le f_i+\beta.
\]
The queried values $\widetilde f_i$ need not be monotone. The following search uses only the monotonicity of $f_i$ and the error bound (\Cref{fig:portal-search}).

For $b\le i$, the difference $f_b-f_i=\delta(u,P[b])+\delta_P(P[b],P[i])-\delta(u,P[i])$ is the extra length of entering $P$ at $P[b]$ on the way to $P[i]$. Hence $P[b]$ satisfies \Cref{lem:portal} for every $x=P[i]$ with $i\ge b$ and $f_b-f_i\le\eps\hat\Delta$, so the thresholds below are evenly spaced in the values of $f$.

Let $h=\eps\hat\Delta/2$. For every threshold
\[
    t_j=-2\hat\Delta+jh,
    \qquad 0\le j\le\ceil{4\hat\Delta/h}+2,
\]
start with indices $\ell=0$ and $b=m+1$, treating their queried values as $+\infty$ and $-\infty$, respectively. While $b-\ell>1$, query the middle index $q$. If $\widetilde f_q>t_j$, set $\ell=q$. Otherwise set $b=q$. At termination, add $P[b]$ if $b\le m$. Include $P[1]$ and $P[m]$ as well.
\begin{figure}[t]
    \centering
    \begin{tikzpicture}[x=0.55cm,y=1cm,>=stealth]
        \foreach \t in {0.2,0.9,1.6,3.0,3.7}
            \draw[gray!45,densely dashed] (0.3,\t) -- (14.7,\t);
        \draw[black!80,thick,dashed] (0.3,2.3) -- (14.7,2.3) node[right] {$t_j$};
        \draw[decorate,decoration={brace,amplitude=3pt}]
            (-0.3,2.3) -- (-0.3,3.0) node[midway,left=3pt] {$h$};

        \draw[->] (0,-0.35) -- (15.4,-0.35) node[right] {$i$};
        \draw[->] (0,-0.35) -- (0,4.3);

        \foreach[count=\i] \f/\e in {3.6/0.1,3.5/0.4,3.1/0.05,3.0/0.35,2.6/0.3,2.0/0.05,1.9/0.4,
                                    1.9/0.1,1.4/0.35,1.0/0.0,0.9/0.4,0.5/0.2,0.4/0.05,0.1/0.3} {
            \fill[gray!25] (\i-0.18,\f) rectangle (\i+0.18,\f+0.45);
            \fill (\i,\f) circle (1.6pt);
            \draw[red!75!black,thick] (\i-0.12,\f+\e-0.07) -- (\i+0.12,\f+\e+0.07)
                (\i-0.12,\f+\e+0.07) -- (\i+0.12,\f+\e-0.07);
        }

        \foreach \i/\f in {1/3.6,5/2.6,6/2.0,10/1.0,12/0.5,14/0.1} {
            \draw[blue!70!black,thick] (\i,\f) circle (3.5pt);
            \draw[blue!70!black,thick] (\i,-0.45) -- (\i,-0.25);
        }
        \node[below,blue!70!black] at (6,-0.45) {$b$};
        \node[below] at (9,-0.45) {$i$};
        \draw[thick] (9,-0.45) -- (9,-0.25);

        \draw[densely dotted] (6,2.0) -- (9.45,2.0);
        \draw[decorate,decoration={brace,mirror,amplitude=3pt}]
            (9.45,1.4) -- (9.45,2.55) node[midway,right=3pt] {$h+\beta$};
    \end{tikzpicture}
    \caption{The values $f_i$ (dots), the bands $[f_i,f_i+\beta]$, the oracle-based values $\widetilde f_i$ (crosses), and the selected in-portals (circled). The search for the threshold $t_j$ returns $P[b]$, which serves the index $i$, since $b\le i$ and $f_b-f_i\le h+\beta$.}
    \label{fig:portal-search}
\end{figure}
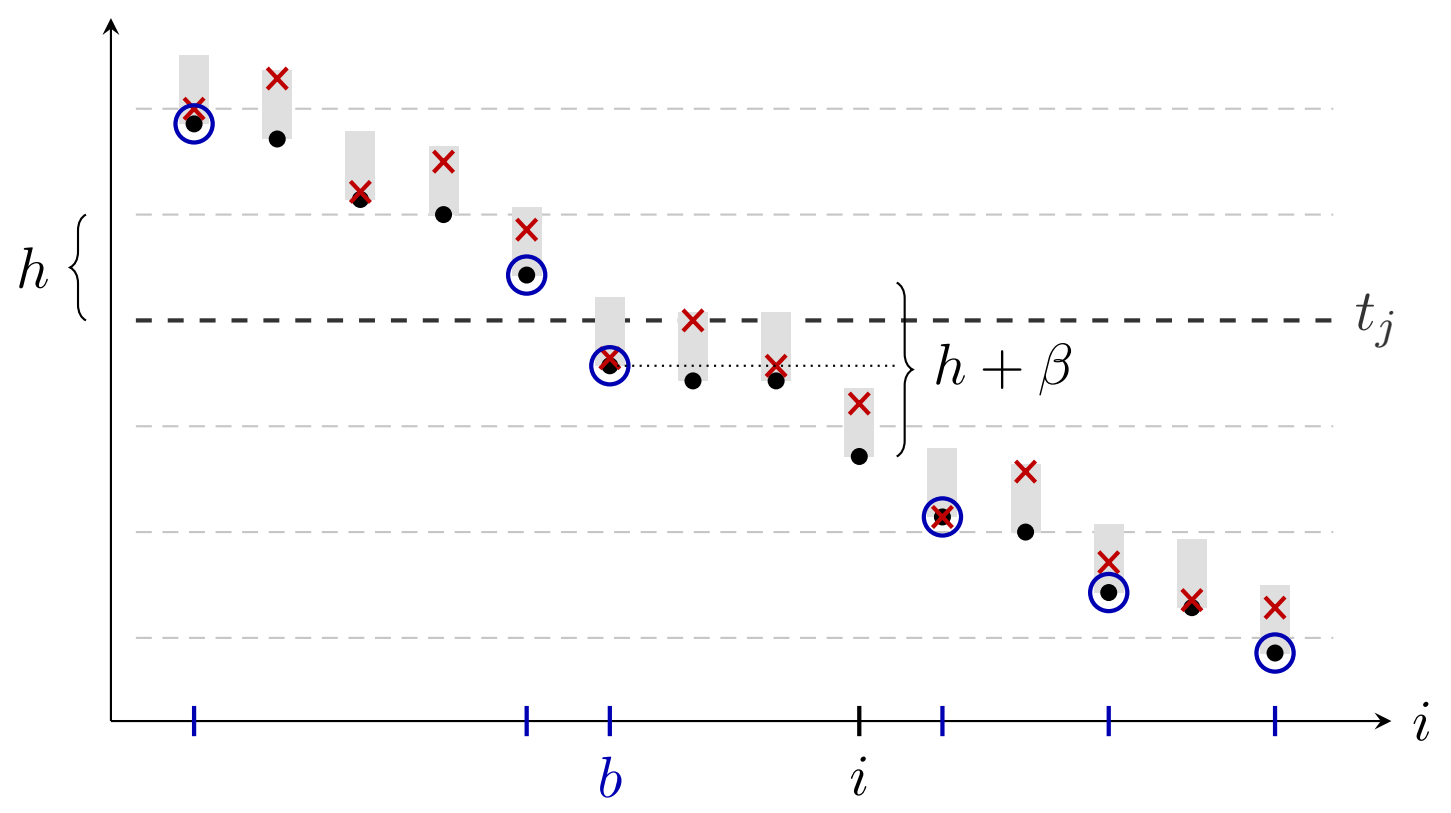

The search maintains $\widetilde f_\ell>t_j$ and $\widetilde f_b\le t_j$ and ends with $\ell=b-1$, so $f_b\le\widetilde f_b\le t_j$ and $f_{b-1}\ge\widetilde f_{b-1}-\beta>t_j-\beta$.

Fix an index $i$. The threshold grid contains a $t_j$ with
\[
    f_i+\beta<t_j\le f_i+\beta+h.
\]
For this search the returned boundary has $b\le i$: otherwise monotonicity would imply $f_i\ge f_{b-1}>t_j-\beta$, a contradiction. Thus the portal $s=P[b]$ satisfies
\begin{align*}
    \delta(u,s)+\delta_P(s,P[i])
    &= f_b+L_i\\
    &\le f_i+\beta+h+L_i
     \le\delta(u,P[i])+\eps\hat\Delta.
\end{align*}
There are $O(1/\eps)$ thresholds, each using $O(\log n)$ probes of cost $O(\log n+Q)$. Store the oracle answer for each selected portal.

\paragraph{Out-portals and total time.}
Apply the same procedure to $P$ in reverse order in $G^R$, querying $\adist(x,u)$ in place of $\adist(u,x)$. The reverse prefix length of $x$ is $\delta(r,z)-\delta(r,x)$, so the same tree preprocessing suffices, and no second oracle is needed for this step. This gives the out-portal guarantee in \Cref{lem:portal}.

There are at most three paths per separator set. Consequently the $O(n\log n)$ relevant incidences take
\[
    O\!\left(n\eps^{-1}\log^2 n\,
        (\log n+\eps^{-1}+\log\log(n/\eps))\right)
\]
time, which is subsumed by the shared oracle's preprocessing bound.
\end{proof}

\end{document}